%% file: paper.tex
\documentclass[
aps, prd, reprint,
10pt, notitlepage, a4paper,
floats, floatfix,
amsmath, amssymb, amsfonts,
superscriptaddress,
showpacs, showkeys,
nofootinbib,
longbibliography,
]{revtex4-1}

\usepackage{graphicx} % include figures
\usepackage{grffile}
\usepackage[usenames,dvipsnames]{xcolor}
\usepackage{xspace} % Sensible space treatment at end of simple macros
\usepackage{bm} % bold math
\usepackage{amsmath,amsfonts,amssymb,amsthm}
\usepackage[utf8]{inputenc} % for some references from inspirehep.net

\usepackage[normalem]{ulem}

\xdefinecolor{mylinkcolor}{rgb}{0,0,0.5}
\usepackage[
	bookmarksnumbered, bookmarksopen, bookmarksopenlevel=2,
	breaklinks=true,
	colorlinks=true, filecolor=mylinkcolor, citecolor=mylinkcolor,
	linkcolor=mylinkcolor, urlcolor=mylinkcolor, menucolor=mylinkcolor,
]{hyperref}

\def\vct#1{{\bm{#1}}}

\def\nl{\\ & \quad}

\DeclareMathOperator{\Order}{\mathcal{O}}

\allowdisplaybreaks

\newcommand{\BB}{SEOB$_\text{TS}$}
\newcommand{\BD}{SEOB$_\text{TM}^{r_c}$}
\newcommand{\simp}{SEOB$_\text{TM}$}
\newcommand{\DN}{SEOB$_\text{TM}^{r_c,\text{align}}$}

\newcommand{\AEI}{\affiliation{Max Planck Institute for Gravitational Physics (Albert Einstein Institute), Am M\"uhlenberg 1, Potsdam 14476, Germany}}
\newcommand{\Maryland}{\affiliation{Department of Physics, University of Maryland, College Park, MD 20742, USA}}

\begin{document}

\title{Fourth post-Newtonian effective-one-body Hamiltonians with generic spins}

\author{Mohammed Khalil}\email{mohammed.khalil@aei.mpg.de}\AEI\Maryland
\author{Jan Steinhoff}\email{jan.steinhoff@aei.mpg.de}\AEI
\author{Justin Vines}\email{justin.vines@aei.mpg.de}\AEI
\author{Alessandra Buonanno}\email{alessandra.buonanno@aei.mpg.de}\AEI\Maryland

\begin{abstract}
  In a compact binary coalescence, the spins of the compact objects
  can have a significant effect on the orbital motion and
  gravitational-wave (GW) emission.  For generic spin orientations,
  the orbital plane precesses, leading to characteristic modulations
  of the GW signal. The observation of precession effects is 
 crucial to discriminate among different binary formation scenarios, 
and to carry out precise tests of General Relativity.
Here, we work toward an improved description of spin effects in binary inspirals,
  within the effective-one-body (EOB) formalism, which is commonly used to build 
waveform models for LIGO and Virgo data analysis. We
  derive EOB Hamiltonians including the complete fourth post-Newtonian
  (4PN) conservative dynamics, which is the current state of the art.
  We place no restrictions on the spin orientations or magnitudes, 
  or on the type of compact object (e.g., black hole or neutron star),
  and we produce the first generic-spin EOB Hamiltonians complete at
  4PN order.  We consider multiple spinning EOB Hamiltonians, which are
  more or less direct extensions of the varieties found in previous
  literature, and we suggest another simplified variant.  Finally, we
  compare the circular-orbit, aligned-spin binding-energy functions
  derived from the EOB Hamiltonians to numerical-relativity
  simulations of the late inspiral. While finding that all proposed Hamiltonians 
  perform reasonably well, we point out some interesting differences, which 
could guide the selection of a simpler, and thus faster-to-evolve EOB Hamiltonian to be used 
in future LIGO and Virgo inference studies.
\end{abstract}

\maketitle

\section{Introduction}

The observation of gravitational waves (GWs) from coalescing binaries
\cite{Abbott:2016blz,TheLIGOScientific:2016pea,TheLIGOScientific:2017qsa,LIGOScientific:2018mvr}
using a continually improving network of GW
detectors~\cite{TheLIGOScientific:2014jea,TheVirgo:2014hva,Aso:2013eba,LIGOIndia}
is a milestone in fundamental physics and astrophysics. As 
the detectors increase their sensitivity, we will observe more events, with 
larger signal-to-noise ratios, spanning 
a larger region of the parameter space. Thus, to faithfully recover 
the sources' properties, it is important to 
improve the accuracy of models of waveforms from binaries of compact objects (black holes and/or neutron
stars) on generic orbits and with generic spin orientations.
In the
generic case, the orbital plane and the objects' spins precess about
the direction of the system's total angular momentum, leading to
modulations of the GW signal which are a smoking-gun signature of the
dynamical influence of the spins. Including such precession
effects in GW template models, as opposed to restricting to the
simpler aligned-spin case with no precession, is important for more
inclusive GW searches, more accurate inference studies and tests 
of General Relativity.

The effective-one-body (EOB) framework~\cite{Buonanno:1998gg,Buonanno:2000ef} aims at providing a synergy between multiple analytical approximations and numerical-relativity (NR) simulations of relativistic inspiraling binaries.  
The core ingredient of the EOB approach is the EOB Hamiltonian, a canonical Hamiltonian describing the binary's (conservative) orbital dynamics, which both (i) agrees, in its post-Newtonian (PN) expansion,\footnote{Recently, also the post-Minkowskian (weak-field) approximation for unbound orbits is considered; see below.} with known results for arbitrary mass ratios from PN calculations (in the weak-field and low-speed regime), and (ii) becomes, in the extreme-mass-ratio limit, an exact Hamiltonian for a test (or probe) particle in an exact black-hole spacetime, valid for arbitrary separations and speeds.  The EOB Hamiltonian is naturally expressed as a deformation of the zero-mass-ratio test-particle Hamiltonian, with the deformation determined by finite-mass-ratio results from the PN approximation.   For example, the original (nonspinning) EOB Hamiltonian \cite{Buonanno:1998gg} becomes, as the mass ratio goes to zero, the exact Hamiltonian for a test mass undergoing geodesic motion in a Schwarzschild (nonspinning black hole) spacetime.  

In generalizing to spinning black holes, the first natural replacement
for the Schwarzschild-geodesic Hamiltonian is the Hamiltonian for
geodesic (test-mass) motion in an exact Kerr (spinning black hole)
spacetime.\footnote{With the aim of building a first
  inspiral-merger-ringdown waveform model for generic spins,
  Ref.~\cite{Buonanno:2005xu} employed a spinning EOB Hamiltonian
  built by adding to the Schwarzschild-geodesic EOB Hamiltonian the
  PN-expanded spin Hamiltonian.}  A spinning EOB (SEOB) Hamiltonian
incorporating the Kerr-geodesic limit was first constructed in
Ref.~\cite{Damour:2001tu} including leading-order (LO) spin-orbit and
LO spin-squared effects in the PN expansion. This was later extended
to the next-to-leading (NLO) \cite{Damour:2008qf} and next-to-NLO
(NNLO) \cite{Nagar:2011fx} spin-orbit levels, and to the NLO
spin-squared level for aligned spins \cite{Balmelli:2013zna,
  Balmelli:2015lva} and then for generic (precessing) spins
\cite{Balmelli:2015zsa}.  The Kerr-geodesic-based approach for aligned
spins has been further developed in 
Refs.~\cite{Pan:2009wj,Damour:2014sva,Nagar:2015xqa,Nagar:2017jdw,Akcay:2018yyh,Nagar:2018zoe,Nagar:2018gnk}, e.g.,
by including matter effects (for neutron stars) and calibration to NR simulations.  A second category of SEOB
Hamiltonians is based on the Hamiltonian for a \emph{spinning} test-body
(test spin) in a Kerr background \cite{Barausse:2009aa,Vines:2016unv}, first
developed with NLO \cite{Barausse:2009xi} and then NNLO
\cite{Barausse:2011ys} spin-orbit terms and with LO spin-squared terms. 
Such Hamiltonians have always been applicable for generic (precessing) spins. They have 
been generalized to include tidal effects in Refs.~\cite{Hinderer:2016eia,Steinhoff:2016rfi}, 
they have been used for studies of extreme-mass-ratio binaries in Ref.~\cite{Yunes:2010zj} and 
periastron advance in Ref.~\cite{Hinderer:2013uwa}, and they have been refined and calibrated to NR simulations in 
Refs.~\cite{Taracchini:2012ig,Taracchini:2013rva,Pan:2013rra,Bohe:2016gbl,Babak:2016tgq,Cotesta:2018fcv}. 
EOB Hamiltonians have also been constructed to include information from gravitational self-force calculations
\cite{Damour:2009sm, Akcay:2012ea, Barausse:2011dq,
  Antonelli:2019fmq} (for extreme mass ratios) and from the post-Minkowskian approximation \cite{Damour:2016gwp, Damour:2017zjx,
  Antonelli:2019ytb} (assuming weak fields but allowing arbitrary speeds).  A recent comparison of
various SEOB waveform models is given in Ref.~\cite{Rettegno:2019tzh}. Waveform models constructed with the 
SEOB Hamiltonians based on a \emph{spinning} test-body in a Kerr background~\cite{Taracchini:2013rva,Pan:2013rra,
Bohe:2016gbl,Babak:2016tgq,Cotesta:2018fcv} have been employed in template banks of LIGO and Virgo, 
and inference studies of binary black holes~\cite{Abbott:2016blz,TheLIGOScientific:2016pea,LIGOScientific:2018mvr}. 
For parameter estimation of binary neutron stars both classes of SEOB Hamiltonians have been employed in Ref.~\cite{LIGOScientific:2018mvr}.

The goal of the present paper is to construct SEOB Hamiltonians for
compact binaries (black holes or neutron stars) that include all known
PN results to 4PN order for generic orbits and spin orientations.
Beyond the up-to-NNLO spin-orbit and spin-squared contributions, the
4PN level includes also the LO cubic and quartic in spin terms.
Previous work is not complete to 4PN order for generic spins
\cite{Barausse:2009xi,Barausse:2011ys,Balmelli:2015zsa} or complete to
4PN but valid for aligned spins only
\cite{Damour:2014sva,Akcay:2018yyh,Nagar:2018plt,Nagar:2018zoe}.  We
construct three SEOB Hamiltonians in this paper: (i) a Hamiltonian
based on Ref.~\cite{Balmelli:2015zsa}, which uses the idea of
``centrifugal radius'' $r_c$ \cite{Damour:2014sva}, while recovering
the Kerr-geodesic limit; (ii) a simplified version of the Hamiltonian
from Ref.~\cite{Balmelli:2015zsa} that does not use a centrifugal
radius and has a different factorization for the PN spin corrections,
similarly recovering the Kerr-geodesic limit; and (iii) one Hamiltonian
following Refs.~\cite{Barausse:2009aa, Barausse:2009xi,
  Barausse:2011ys} which recovers the dynamics of a spinning test-body
in the Kerr spacetime in the small-mass-ratio limit (see Table
\ref{tab:SEOBdefs} for a summary of the differences between these
Hamiltonians).  As we wish to somewhat fairly compare different
treatments of spin effects in the EOB formalism, we have modified some
details of the original proposals of
Refs.~\cite{Balmelli:2015zsa,Barausse:2011ys} such that all the
Hamiltonians agree in the zero-spin limit.  We compare the
aligned-spin circular-orbit binding energy functions from the different
Hamiltonians with NR simulations and with the aligned-spin Hamiltonian
from Refs.~\cite{Nagar:2018plt,Nagar:2018zoe,Damour:2014sva}. This
enables one to assess compromises between accuracy and simplicity of
the SEOB Hamiltonians.

The paper is organized as follows.
In Sec.~\ref{spinEOB} we provide an overview of SEOB Hamiltonians and their construction.
Sections \ref{EOB4PN} and \ref{sec:BB} present the ans{\"a}tze of the SEOB Hamiltonians, with explicit results matched to 4PN in Appendix~\ref{app:coeffs}.
We then compare the aligned-spin circular-orbit binding energy of the Hamiltonians against NR in Sec.~\ref{compare}.
Our conclusions are given in Sec.~\ref{conclude}.
Appendix~\ref{app:BDcorrect} corrects an omission at NLO S$^2$ in the Hamiltonian of Ref.~\cite{Balmelli:2013zna}.

\begin{table*}[t]
\caption{The SEOB Hamiltonians used in this paper and the differences between them. All Hamiltonians include complete 4PN results for generic spins and compact objects, except for the last Hamiltonian, which is for aligned spins. }
\begin{ruledtabular}
\begin{tabular}{llc}
SEOB        &  definition & references \\ 
	\hline 
{\BB} & based on the Hamiltonian for a test spin (TS) in Kerr spacetime & \cite{Barausse:2009xi,Barausse:2011ys} \\
{\BD} & based on the Hamiltonian for a test mass (TM) in Kerr spacetime; it uses the centrifugal radius $r_c$ & \cite{Balmelli:2015zsa} \\
{\simp} & simplified version of \BD; it does not use $r_c$; it uses different factorization for spin corrections & [this paper] \\
{\DN} & similar to {\BD}, but for aligned spins and includes S$^2$ and S$^4$ corrections, differently & \cite{Damour:2014sva,Nagar:2018plt,Nagar:2018zoe}\\
\end{tabular}
\end{ruledtabular}
\label{tab:SEOBdefs}
\end{table*}

\subsection*{Notation}

We use geometric units such that the speed of light $c$ and the Newton constant $G$ are equal to 1.
We utilize various combinations of the masses $m_1$, $m_2$ of the binary's components,
\begin{gather}\label{massmap}
M= m_1 + m_2, \quad \mu = \frac{m_1m_2}{M}, \quad \nu = \frac{\mu}{M}, \nonumber\\
q = \frac{m_1}{m_2}, \quad X_1=\frac{m_1}{M}, \quad X_2=\frac{m_2}{M}.
\end{gather}
For the spins $\bm{S}_1$, $\bm{S}_2$, we define the dimensionless versions
\begin{gather}
\bm{\chi}_1 = \frac{\bm{a}_1}{m_1} = \frac{\bm{S}_1}{m_1^2}, \qquad 
\bm{\chi}_2 = \frac{\bm{a}_2}{m_2} = \frac{\bm{S}_2}{m_2^2},
\end{gather}
along with the intermediate $\vct a_1$, $\vct a_2$.
The relative position and momentum are denoted by $\vct{r}$ and $\vct{p}$, respectively.
Using an implicit Euclidean background, it holds
\begin{equation}
\vct p^2 = p_r^2 + \frac{L^2}{r^2}, \quad
p_r= \bm{n}\cdot\bm{p}, \quad
\bm{L}=\bm{r}\times\bm{p},
\end{equation}
where $\bm{n}=\bm{r}/r$ with $r=|\vct r|$, and $\vct L$ is the orbital angular momentum with magnitude $L$.
For convenience, we also introduce rescaled dimensionless variables,
\begin{equation}
\label{reducedVar}
\hat{\vct r}=\frac{\vct r}{M}, \quad \hat{\vct p}=\frac{\vct p}{\mu}, \quad \hat{H}=\frac{H}{\mu}, \quad 
\hat{\vct L}=\frac{\vct L}{M\mu}, \quad \hat{\vct a}=\frac{\vct a}{M},
\end{equation}
and similarly for the magnitudes $\hat r=|\hat{\vct r}|$, etc.; here, $H$ is any of several Hamiltonians encountered below, and $\vct a=\vct S_\text{Kerr}/M$ is the rescaled spin of an effective Kerr black hole.

\section{Spinning effective-one-body Hamiltonians}
\label{spinEOB}

In this section, we give an overview of spinning EOB Hamiltonians and their construction \cite{Buonanno:1998gg, Damour:2001tu, Damour:2008qf, Nagar:2011fx, Barausse:2009aa, Barausse:2009xi, Barausse:2011ys, Balmelli:2013zna, Damour:2014sva, Balmelli:2015lva, Balmelli:2015zsa, Nagar:2018zoe, Nagar:2018plt}, on which current EOB waveform models are built \cite{Buonanno:2000ef, Damour:2008gu, Taracchini:2012ig, Taracchini:2013rva, Pan:2013rra, Bohe:2016gbl, Babak:2016tgq, Nagar:2015xqa, Nagar:2017jdw, Nagar:2018gnk, Nagar:2018zoe}.
The EOB Hamiltonians are constructed such that (i) they describe geodesic motion in Kerr spacetime in the limit of vanishing mass ratio and that (ii) they agree (up to a canonical transformation) with a PN approximate Hamiltonian describing the conservative binary motion up to a certain order (here the 4PN order \cite{Levi:2016ofk}).
A certain class of EOB Hamiltonians \cite{Barausse:2009xi, Barausse:2009aa, Barausse:2011ys} also incorporates the (nongeodesic) motion of spinning test particles in Kerr spacetime in the small mass-ratio limit, as described by the Matthisson-Papapetrou-Dixon equations \cite{Mathisson:1937zz, Mathisson:2010, Papapetrou:1951pa, Corinaldesi:1951pb, Dixon:1979}.

We consider a spinning binary in the center-of-mass frame.
The orbital dynamics is described by the relative separation $\vct{r}$ and linear momentum $\vct{p}$ vectors, and the internal dynamics is assumed to be captured by the spins $\vct{S}_1$ and $\vct{S}_2$ of each body.
The Poisson brackets between these dynamical variables are the standard ones,
\begin{subequations}
\begin{align}
  \{ r^i, p_j \} &= \delta_{ij} , \\
  \{ S_1^i, S_1^j \} &= \epsilon_{ijk} S_1^k , \\
  \{ S_2^i, S_2^j \} &= \epsilon_{ijk} S_2^k ,
\end{align}
\end{subequations}
with all others vanishing.  The dynamics on phase space is generated by a Hamiltonian function $H(\vct r,\vct p,\vct S_1,\vct S_2)$.
The equation of motion of a generic phase-space function $A$ reads
\begin{equation}
  \frac{d A}{d t} = \{ A, H \} + \frac{\partial A}{\partial t} .
\end{equation}
Here the Hamiltonian is either the PN $H^\text{PN}$ or the EOB $H^\text{EOB}$ one.
The EOB Hamiltonian $H^\text{EOB}$ itself is given in terms of another Hamiltonian, the effective Hamiltonian $H^\text{eff}$, via the energy map,
\begin{equation}
H^\text{EOB} = M \sqrt{1 + 2\nu \left(\frac{H^\text{eff}}{\mu} - 1\right)}\,.
\end{equation}
The utility of this energy map was demonstrated, e.g., in Refs.~\cite{Buonanno:1998gg, Damour:2000we, Damour:2016gwp}.
For instance, if for $H^\text{eff}$ one just takes the Hamiltonian of geodesics in Schwarzschild spacetime, then $H^\text{EOB}$ correctly describes both the 1PN and first post-Minkowskian dynamics \cite{Buonanno:1998gg, Damour:2016gwp}.

\subsection{The effective Hamiltonian}
The central idea of the EOB Hamiltonian is to combine the dynamics in the test-body limit (with no restriction on the speed or field strength) with the PN dynamics (not restricted in the mass ratio).
In this way, one might overcome some of the limitations of the individual approximations.
This can be achieved by making an ansatz for $H^\text{eff}$ as a deformation of the test-body-limit Hamiltonian (deforming it such that PN results are recovered), which is the purpose of this section.
Note that in the test-body limit $H^\text{EOB} \approx H^\text{eff} + \text{const}$.

Let us review the Hamiltonian of a spinning test-body in Kerr spacetime \cite{Barausse:2009aa, Vines:2016unv}.
One can easily specialize this to the nonspinning (geodesic) case, which is the basis of some SEOB models.
These test-body Hamiltonians are the basis for all SEOB models.
The (inverse) Kerr metric $g_\text{Kerr}^{\mu\nu}$ in Boyer-Lindquist coordinates $(x^\mu) = (t, r, \theta, \phi)$ is given by the line element
\begin{align}
\label{KerrH}
-d\tau^2 &=
g_\text{Kerr}^{\mu\nu} \partial_\mu \partial_\nu
\nonumber\\
&=- \frac{\Lambda}{\Delta \Sigma} \partial_t^2+\frac{\Delta}{\Sigma}\partial_r^2+\frac{1}{\Sigma}\partial_\theta^2 \nonumber\\
&\quad +\frac{\Sigma-2Mr}{\Sigma\Delta\sin^2\theta}\partial_\phi^2-\frac{4Mra}{\Sigma\Delta}\partial_t\partial_\phi,
\end{align}
where $M$ is the mass of the black hole, $\sigma = M a$ is its spin, and
\begin{subequations}
\begin{gather}
\Sigma \equiv r^2 + a^2 \cos^2\theta , \qquad 
\Delta \equiv r^2 - 2 M r + a^2 , \\
\Lambda \equiv (r^2+a^2)^2-a^2 \Delta \sin^2\theta .
\end{gather}
\end{subequations}
The Hamiltonian of a spinning test-body $H^\text{Kerr}$ can be obtained as a solution of the mass-shell constraint (see, e.g., Ref.~\cite{Vines:2016unv})
\begin{equation}
  \begin{split}
    -\mu^2 &= g_\text{Kerr}^{\mu\nu}\left( p_\mu - \frac{1}{2} \omega_{\mu ab} S_*^{ab} \right) \left(p_\nu - \frac{1}{2} \omega_{\mu ab} S_*^{ab} \right) \nl
    + \Order(S_*^2) ,
  \end{split}
\end{equation}
where $p_\mu=(-H^\text{Kerr},p_r,p_\theta,p_\phi)$, $\mu$ is the mass of the test-body, $S_*^{ab}=-S_*^{ab}$ is its spin tensor in a local Lorentz frame ($e_a{}^\mu e^{a\nu} = g_\text{Kerr}^{\mu\nu}$), $ \omega_{\mu ab} = e_{b \nu} \nabla_\mu e_a{}^\nu$ are the Ricci rotation coefficients, and $\nabla_\mu$ is the covariant derivative.
The canonical spin vector of the test-body $\vct{S}_*$ is given by $S^i_* = \frac{1}{2} \epsilon^{ijk} S_*^{jk}$ and the components $S_{0i}$ are fixed by the supplementary condition $S^{ab} (e_b{}^\mu p_\mu + \mu^2 \delta^0_b) = 0 + \Order(S_*^2)$, all in the local frame.

Let us split $H^\text{Kerr}$ into a part dependent on the test-spin $\vct{S}_*$ and the remaining $S_*$-independent terms into parts even and odd in the Kerr spin $a$,
\begin{equation}
H^\text{Kerr} = H^\text{Kerr}_\text{even} + H^\text{Kerr}_\text{odd} + H^\text{Kerr}_{S_*}.
\end{equation}
Following the procedure outlined above, and choosing the local frame from Ref.~\cite{Barausse:2009xi}, this leads to
\begin{subequations}
\begin{align}
  H^\text{Kerr}_\text{even} &= \alpha^\text{Kerr} \sqrt{\mu^2 + \gamma_\text{Kerr}^{\phi\phi} p_\phi^2 + \gamma_\text{Kerr}^{rr} p_r^2 + \gamma_\text{Kerr}^{\theta\theta} p_\theta^2} , \\
  H^\text{Kerr}_\text{odd} &= \beta^\text{Kerr} p_\phi \\
  H^\text{Kerr}_{S_*} &= \left[\bm{F}_t+\left(\beta^\text{Kerr}+\frac{\alpha^\text{Kerr} \gamma_\text{Kerr}^{\phi\phi} p_\phi}{\sqrt{q^\text{Kerr}}}\right)\bm{F}_\phi\right]\cdot \bm{S}_*\nonumber\\
&\quad +\frac{\alpha^\text{Kerr}
}{\sqrt{q^\text{Kerr}}}\left(\gamma_\text{Kerr}^{rr}p_r \bm{F}_r+
\gamma_\text{Kerr}^{\theta\theta}p_\theta \bm{F}_\theta\right)\cdot \bm{S}_* \nonumber\\
&\quad  + \Order(S_*^2) , \label{Hodd}
\end{align}
\end{subequations}
with
\begin{subequations}
\begin{align}
  \alpha^\text{Kerr} &= \frac{1}{\sqrt{- g_\text{Kerr}^{tt}}} = \sqrt{\frac{\Delta \Sigma}{\Lambda}}, \\
  \beta^\text{Kerr} &= \frac{g_\text{Kerr}^{t\phi}}{g_\text{Kerr}^{tt}} = \frac{2 a M r}{\Lambda}, \\
  \gamma_\text{Kerr}^{\phi\phi} &= g_\text{Kerr}^{\phi\phi} - \frac{g_\text{Kerr}^{t\phi} g_\text{Kerr}^{t\phi}}{g_\text{Kerr}^{tt}} = \frac{\Sigma}{\Lambda \sin^2\theta} , \\
  \gamma_\text{Kerr}^{rr} &= g_\text{Kerr}^{rr} = \frac{\Delta}{\Sigma} , \\
  \gamma_\text{Kerr}^{\theta\theta} &= g_\text{Kerr}^{\theta\theta} = \frac{1}{\Sigma} , \\
  \sqrt{q^\text{Kerr}} &= \frac{H^\text{Kerr}_\text{even}}{\alpha^\text{Kerr}},
\end{align}
\end{subequations}
and with explicit expressions for the fictitious gravito-magnetic (frame-dragging) force interacting with the test-spin $\vct{S}_*$  given in Ref.~\cite{Hinderer:2013uwa} in terms of the vectors $\vct{F}_\mu$ (reproduced here in Sec.~\ref{sec:BB}).
A simplified version of this Hamiltonian for aligned spins and motion in the equatorial plane can be found in Ref.~\cite{Bini:2015xua}.
Simplifications for the generic-spin case are possible by making a different choice for the local frame which may simplify the Ricci rotation coefficients, see, e.g., Appendix C of Ref.~\cite{Vines:2016unv}.

The Hamiltonian above is written in terms of components instead of vectors, which is a disadvantage for some purposes.
Following Ref.~\cite{Balmelli:2015zsa}, we transform to a 3-vector notation (with an implicit flat Euclidean background) by treating $(r,\theta,\phi)$ as spherical coordinates, with $\vct{r} = (x,y,z)=r(\sin\theta\cos\phi,\sin\theta\sin\phi,\cos\theta)$ and $\vct{a} = (0,0,a)$.
This is accompanied by a transformation of the momenta $p_r$, $p_\theta$, $p_\phi$ to the new momenta $\vct{p}$,
\begin{subequations}
\begin{gather}
  p_r = \bm{n}\cdot\bm{p}, \quad p_\phi = L_z = (\vct{r} \times \vct{p})_z \\
   \frac{p_\theta^2}{r^2} = \vct{p}^2 - p_r^2 - \frac{p_\phi^2}{r^2 \sin^2 \theta}
\end{gather}
\end{subequations}
which makes it an overall canonical transformation.
Noting that $a^2p_\phi^2/r^2 = (\bm{n} \times \bm{p} \cdot \bm{a})^2 $, $a \cos \theta = \vct{n \cdot a}$, and $a^2 \sin^2 \theta = a^2 - (\vct{n \cdot a})^2$, this results in the even-in-$a$ Hamiltonian
\begin{subequations}
\begin{align}
H^\text{Kerr}_\text{even} =
\bigg[ &
A^\text{Kerr}
\Big( \mu^2 + B_p^\text{Kerr} \bm{p}^2  + B_{np}^\text{Kerr} (\bm{n}\cdot\bm{p})^2   \nonumber\\
& + B_{npa}^\text{Kerr} (\bm{n} \times \bm{p} \cdot \bm{a})^2
\Big)
\bigg]^{1/2},
\end{align}
\end{subequations}
with
\begin{subequations}
\begin{align}
A^\text{Kerr} &= (\alpha^\text{Kerr})^2 = \frac{\Delta \Sigma}{\Lambda} , \label{AKerr} \\
B_{p}^\text{Kerr} &= r^2 \gamma_\text{Kerr}^{\theta\theta} = \frac{r^2}{\Sigma} , \\
B_{np}^\text{Kerr} &= \gamma_\text{Kerr}^{rr} - r^2 \gamma_\text{Kerr}^{\theta\theta} =  \frac{r^2}{\Sigma} \left[\frac{\Delta}{r^2} - 1\right] , \\
  B_{npa}^\text{Kerr}  &= \frac{r^2}{a^2} \left[ \gamma_\text{Kerr}^{\phi\phi} - \frac{\gamma_\text{Kerr}^{\theta\theta}}{\sin^2 \theta} \right]
= -\frac{r^2}{\Sigma \Lambda} (\Sigma + 2 M r) ,
\end{align}
\end{subequations}
and
\begin{subequations}
\begin{gather}
\Sigma = r^2 + (\vct{n} \cdot \vct{a})^2 , \qquad 
\Delta = r^2 - 2 M r + a^2 , \\
\Lambda = (r^2 + a^2)^2 - \Delta a^2 + \Delta (\vct{n} \cdot \vct{a})^2 .
\end{gather}
\end{subequations}
Similarly, the odd-in-$a$ part reads
\begin{equation}
  H_\text{odd}^\text{Kerr} = \beta^\text{Kerr} p_\phi = \frac{2 M r}{\Lambda} \bm{L} \cdot \bm{a} .
\end{equation}

We now have all ingredients in order to discuss how an ansatz for the effective Hamiltonian can be built.
In general, one takes the effective Hamiltonian to be a deformation of the Kerr Hamiltonian (the deformation parameter being the symmetric mass ratio $\nu$), either for a test spin or a test mass.
While all EOB models agree on the identification of the masses between the test-body and comparable mass case ($M = m_1 + m_2$, $\mu = m_1 m_2 / M$), different choices are made for mapping the spins $\vct{a}$ and $\vct{S}_*$ to $\vct{S}_1$ and $\vct{S}_2$.
Let us consider a simple explicit example.
We could write the even-in-$a$ part of the effective Hamiltonian as
\begin{align}\label{Heffansatz}
H^\text{eff}_\text{even} &=
\bigg[ 
A
\Big( \mu^2 + B_p \bm{p}^2  + B_{np} (\bm{n}\cdot\bm{p})^2   \nonumber\\
&\quad\quad + B_{npa} (\bm{n} \times \bm{p} \cdot \bm{a})^2 + \mu^2Q  
\Big)
\bigg]^{1/2},
\end{align}
where the momentum-independent potentials $A$, $B_p$, $B_{np}$, $B_{npa}$ are the Kerr potentials given above modified by PN corrections (to be determined). The quantity $Q$ is a momentum-dependent potential introduced in Ref.~\cite{Damour:2000we}, which may accommodate PN terms that do not fit into the momentum-independent potentials.
(In cases where $Q$ vanishes, the deformed Hamiltonian can be interpreted as describing geodesic motion in a $\nu$-deformed Kerr metric.)
The mentioned potentials should all be of even order in spin, while terms of odd order in spin should be included via a deformation of $H^\text{Kerr}_\text{odd}$.
More explicit ans{\"a}tze for the PN-corrected SEOB Hamiltonians and their potentials are discussed below.

\subsection{Matching to post-Newtonian  results}
\label{sec:matching}
To fix the potentials in the ansatz for an effective Hamiltonian, one demands that the EOB Hamiltonian $H^\text{EOB}$ agrees with the Hamiltonian in the PN approximation $H^\text{PN}$ up to a canonical transformation.
This will eventually not uniquely fix the potentials, but leave some (gauge) freedom.

Here we use the spinning PN Hamiltonian derived in the framework and gauges introduced in Ref.~\cite{Levi:2015msa}, since it is available to 4PN order in the spinning sector \cite{Levi:2016ofk}.
Broken up into leading order (LO), next-to-LO (NLO), next-to-NLO (NNLO) PN parts and into powers of spin, it reads
\begin{align}
  &H^\text{PN}_\text{spin} = \\
  &\begin{array}{l@{\,}l@{\,}l@{\,}l@{\,}l@{\,}l@{\,}}
     H^\text{LO}_{S} & &+ H^\text{NLO}_{S} & &+ H^\text{NNLO}_{S} & \\ 
     & + H^\text{LO}_{S^2} & &+ H^\text{NLO}_{S^2} & &+ H^\text{NNLO}_{S^2} \\ 
     & & & &+ H^\text{LO}_{S^3} & \\
     & & & & &+ H^\text{LO}_{S^4} \\
     & & & & &\qquad\quad \ddots \\     
\Order(\frac{1}{c^3}) &+ \Order(\frac{1}{c^4}) &+ \Order(\frac{1}{c^5}) &+ \Order(\frac{1}{c^6}) &+ \Order(\frac{1}{c^7}) &+ \Order(\frac{1}{c^8})
  \end{array} \nonumber
\end{align}
where columns correspond to PN orders counted by the inverse of the speed of light $c$ (one PN order is $\Order(c^{-2})$).
Except for the self-spin-squared interactions in $H^\text{NNLO}_{S^2}$ calculated in Ref.~\cite{Levi:2015ixa}, these results have been derived in different frameworks and checked against each other:
$H^\text{LO}_{S}$ in Refs.~\cite{Tulczyjew:1959,Barker:1970zr,Barker:1975ae,D'Eath:1975vw,Barker:OConnell:1979,Damour:1982,Thorne:1984mz,Damour:1988mr},
$H^\text{NLO}_{S}$ in Refs.~\cite{Tagoshi:2000zg,Faye:2006gx,Damour:2007nc,Steinhoff:2008zr,Perrodin:2010dy,Porto:2010tr,Levi:2010zu},
$H^\text{NNLO}_{S}$ in Refs.~\cite{Hartung:2011te,Hartung:2013dza,Levi:2015uxa,Marsat:2012fn,Bohe:2012mr},
$H^\text{LO}_{S^2}$ in Refs.~\cite{Barker:1975ae,D'Eath:1975vw,Barker:OConnell:1979,Poisson:1997ha,Thorne:1984mz},
$H^\text{NLO}_{S^2}$ in Refs.~\cite{Steinhoff:2007mb,Porto:2006bt,Porto:2008tb,Levi:2008nh,Porto:2008jj,Steinhoff:2008ji,Hergt:2010pa,Hergt:2011ik,Bohe:2015ana},
$H^\text{NNLO}_{S^2}$ in Refs.~\cite{Hartung:2011ea,Levi:2011eq,Hartung:2013dza,Levi:2014sba,Levi:2015ixa},
$H^\text{LO}_{S^3}$ in Refs.~\cite{Hergt:2007ha,Hergt:2008jn,Levi:2014gsa,Marsat:2014xea,Vaidya:2014kza},
and $H^\text{LO}_{S^4}$ in Refs.~\cite{Levi:2014gsa,Hergt:2007ha,Hergt:2008jn,Vaidya:2014kza}.
These Hamiltonians are valid for both black holes and neutron stars.
They depend on coefficients ($\tilde{C}_{(\text{ES}^2)}$, $\tilde{C}_{(\text{BS}^3)}$, $\tilde{C}_{(\text{ES}^4)}$) which are the proportionality constants between the spin-induced multipoles (quadrupole, octupole, hexadecapole) and symmetric-tracefree tensors built out of (two, three, four) spin vectors (respectively).
The proportionality constants depend on the type of compact object (and on the equation of state in case of a neutron star); here they are normalized to 0 for black holes (in the original paper \cite{Levi:2016ofk}, they are normalized to 1 and denoted without a tilde; see also Appendix~\ref{app:coeffs}).
This normalization makes sense here since we base the EOB Hamiltonian on a deformation of the Kerr one.
Of course, the PN Hamiltonian $H^\text{PN} = H^\text{PN}_\text{ns} + H^\text{PN}_\text{spin}$ must be supplemented by its nonspinning (ns) part $H^\text{PN}_\text{ns}$, which we only need to 2PN order here in order to construct the canonical transformation of the spin sector;
it can be derived, e.g., from the Lagrangian in Ref.~\cite{Gilmore:2008gq}.
The nonspinning part was derived to 4PN order using independent methods \cite{Damour:2014jta,Bernard:2016wrg,Foffa:2019rdf,Blumlein:2020pog} and partial results at 5PN have already been obtained \cite{Foffa:2019hrb,Blumlein:2019zku,Bini:2019nra}.

The condition that the EOB Hamiltonian $H^\text{EOB}$ must coincide with results for the PN-approximate binary Hamiltonian $H^\text{PN}$ up to a canonical transformation reads
\begin{multline}\label{eq:trafo}
H^\text{EOB} = H^\text{PN} + \lbrace \mathcal{G}, H^\text{PN} \rbrace
+ \frac{1}{2!}\lbrace \mathcal{G},\lbrace \mathcal{G}, H^\text{PN} \rbrace\rbrace \\
+ \frac{1}{3!}\lbrace \mathcal{G},\lbrace \mathcal{G},\lbrace \mathcal{G}, H^\text{PN} \rbrace\rbrace\rbrace + \dots
\end{multline}
where $\mathcal{G}$ is the generating function of the canonical transformation.
If $\mathcal{G}$ is small in the PN approximation, then the series in Eq.~\eqref{eq:trafo} terminates after a finite number of terms at a given PN order.
In practice, one makes a PN-approximate and manifestly rotation invariant ansatz for $\mathcal{G}$ in terms of the canonical variables;
we provide an explicit expression for $\mathcal{G}$ as \texttt{Mathematica} code in the supplementary material.
Equation \eqref{eq:trafo} then leads to constraints on the coefficients in the ansatz for $\mathcal{G}$ and $H^\text{eff}$.
The remaining freedom in the coefficients is a gauge freedom within the EOB formalism.

Let us note some general considerations about how part of this gauge freedom can be fixed in SEOB models.
Since binaries are expected to be on almost circular orbits during their last orbits, it makes sense to fix the gauge freedom of the EOB Hamiltonian such that it simplifies for circular orbits, for which $p_r \equiv \vct{n} \cdot \vct{p} = 0$ \cite{Damour:2000we}.
Taking the ansatz in Eq.~\eqref{Heffansatz} as an example, this means that---using the canonical transformation discussed above---one should transform as many PN terms as possible into a form such that they can be included in the potential $B^\text{Kerr}_{np}$, which drops out of the Hamiltonian for circular orbits.
In the nonspinning case, it is additionally possible to require that the potential $Q$ depends on the momentum only via $p_r$, and this uniquely fixes all EOB gauge freedom \cite{Buonanno:1998gg,Damour:2000we}.
For the example in Eq.~\eqref{Heffansatz}, following the structure of the nonspinning Hamiltonian, it is natural to require that:
(i) the momentum-dependence of $H_\text{odd}^\text{Kerr}$ is expressed in terms of $p_r$ whenever possible \cite{Damour:2008qf},
(ii) $B_{npa} = B_{npa}^\text{Kerr}$ \cite{Balmelli:2015zsa}, and (iii) terms in $Q$ have a power in $p_r$ that is as high as possible.
The last requirement ensures that $Q$ vanishes for circular orbits, as in the nonspinning case.

These considerations still leave some remaining gauge freedom in the spinning case, which we fix such to simplify the EOB Hamiltonian also for \emph{aligned} spins. For example, it is possible to choose PN corrections in the potential $B_p$ such that it only depends on terms of the form $\bm{n}\cdot\bm{S}$ but not $\bm{S} \cdot \bm{S}$.
Any remaining gauge freedom beyond that may be chosen arbitrarily.

\section{Spinning effective-one-body Hamiltonians with test mass}
\label{EOB4PN}

In this section we present different ans{\"a}tze for effective Hamiltonians based on the Kerr geodesic one.
That is, we do not include the Kerr test-spin Hamiltonian $H_{S_*}^\text{Kerr}$ here and instead make an ansatz of the form
\begin{equation}
H^\text{eff} = H^\text{eff}_\text{even} + H^\text{eff}_\text{odd} .
\end{equation}
The explicit lengthy results from the matching at 4PN order against PN results (and fixing of the remaining gauge freedom) are given in Appendix~\ref{app:coeffs}.
We start with an extension of the SEOB Hamiltonian from Ref.~\cite{Balmelli:2015zsa}, which we call {\BD}, to 4PN order, here including spin effects at LO S$^3$, and NNLO S$^2$.
We also extend that Hamiltonian from black holes to generic compact objects, e.g., neutron stars.
We proceed with a simplified version of the {\BD} Hamiltonian to 4PN order that does not make use of the  ``centrifugal radius'' introduced in Ref.~\cite{Damour:2014sva}.
For completeness, we also summarize the {\DN} Hamiltonian from Refs.~\cite{Damour:2014sva,Akcay:2018yyh,Nagar:2018plt,Nagar:2018zoe} which is valid for aligned spins only.
We do not include additional PN terms in the {\DN} Hamiltonian since it is already 4PN complete for generic bodies. For convenience we summarize the Hamiltonians in Table~\ref{tab:SEOBdefs}.

\subsection{Effective-one-body Hamiltonian with test-mass limit and centrifugal radius: {\BD}}
\label{sec:BD}

Reference \cite{Balmelli:2015zsa} was the first to construct an SEOB Hamiltonian with NLO spin-squared terms for generic spin orientations (but omitting a subtle contribution, included here, see Appendix \ref{app:BDcorrect}). Here we extend the Hamiltonian to include NNLO spin-squared and LO spin-cubed terms, and add multipole constants to make it applicable to generic bodies like neutron stars.

For the even-in-spin part of the {\BD} Hamiltonian, we use the ansatz in Eq.~\eqref{Heffansatz},
\begin{align}
\label{Heffansatz2}
{H}^\text{eff}_\text{even} &=
\bigg[ 
A
\Big( \mu^2 + B_p \bm{p}^2  + B_{np} (\bm{n}\cdot\bm{p})^2   \nonumber\\
&\quad\quad + B_{npa} (\bm{n} \times \bm{p} \cdot \bm{a})^2 + \mu^2Q  
\Big)
\bigg]^{1/2},
\end{align}
where the Kerr spin is mapped according to
\begin{equation}\label{BDspinmap}
  \vct{a} = \vct{a}_1 + \vct{a}_2\,. 
\end{equation}
This ensures that the Hamiltonian reproduces leading-order PN results at all even orders in spin \cite{Vines:2016qwa}.
The effective Hamiltonian further uses the ``centrifugal radius'' $r_c$, which was introduced in Ref.~\cite{Damour:2014sva} and is defined such that the Kerr Hamiltonian for aligned spins and equatorial orbits can be written as $H_\text{even}^\text{Kerr} = \sqrt{A^\text{Kerr} \left(\mu^2 + p_\phi^2 / r_c^2 + p_r^2 / B(r)\right)}$, which implies the definition
\begin{equation}
r_c = \sqrt{r^2 + a^2 + \frac{2Ma^2}{r}}\,.
\end{equation}

The centrifugal radius was generalized to generic spin orientations in Ref.~\cite{Balmelli:2015zsa}.
In terms of $r_c$, the Kerr potential $A^\text{Kerr}$ from Eq.~\eqref{AKerr} can be written equivalently as
\begin{equation}
A^\text{Kerr} = \left(1 - \frac{2M}{r_c}\right) \frac{\left(1 + \frac{2M}{r_c}\right)}{\left(1 + \frac{2M}{r}\right)} 
\frac{1 + \frac{(\bm{n}\cdot\bm{a})^2}{r^2}}{1 + \Delta \frac{(\bm{n}\cdot\bm{a})^2}{r^2 r_c^2}}\,.
\end{equation}
In the nonspinning limit, only the first term above remains, which reduces to the Schwarzschild $A$-potential. This is the reason why Ref.~\cite{Balmelli:2015zsa} adds the zero-spin PN corrections to $1 - 2M/r_c$. However, in this paper we intend to investigate spin effects across different Hamiltonian descriptions, so we need to make sure that the nonspinning Hamiltonians are identical.
That is, we need to choose a method for adding zero-spin corrections that can be applied to all four EOB Hamiltonians considered here. 
We simply multiply the Kerr potential $A^\text{Kerr}$ by zero-spin PN corrections denoted $A^0$ below (without performing a Pad\'{e} or $\log$ resummation\footnote{The justification for the Pad\'{e} or $\log$ resummations is that they improve agreement with NR in some models and may hence be seen as an implicit calibration. In this paper, however, we consider EOB Hamiltonians with no calibration to NR, so we try to avoid such resummations, in particular in the nonspinning part.\label{nopade}} of $A^0$).
For the spin-squared corrections, we follow Ref.~\cite{Balmelli:2015zsa} and add spin-squared corrections of the form $\bm{n}\cdot\bm{S}$ to the term $1+(\bm{n}\cdot\bm{a})^2/r^2$, and add corrections of the form $\bm{S}\cdot\bm{S}$ to $1+2M/r_c$, since it has an expansion of the form $1+2M/r-Ma^2/r^3+\dots$.
One employs similar considerations for adding PN corrections to the $B$-potentials in Eq.~\eqref{Heffansatz2}, leading to the following ansatz:
\begin{widetext}
\begin{subequations}
\begin{align}
A &= \left(1-\frac{2M}{r_c}\right) \frac{\left(1 + \frac{2M}{r_c} + A^{SS} + A^{S^4}\right)}{\left(1 + \frac{2M}{r}\right)} 
\frac{\left(1 + \frac{(\bm{n}\cdot\bm{a})^2}{r^2} + A^{nS}\right)}{\left(1 + \Delta \frac{(\bm{n}\cdot\bm{a})^2}{r^2 r_c^2}\right)} A^0(r_c) \,, \\
B_{p} &= \left[1 + \frac{(\bm{n}\cdot\bm{a})^2}{r^2} + B_p^{nS}\right]^{-1}, \\ 
B_{np} &= \frac{1}{1 + \frac{(\bm{n}\cdot\bm{a})^2}{r^2}} \left[ \left(1 - \frac{2M}{r} + \frac{a^2}{r^2}\right) \left(A^0(r_c) \, D^0(r_c) +  B_{np}^{SS} + B_{np}^{nS} \right)  - 1\right], \\
B_{npa} &= B_{npa}^\text{Kerr}\,, \\
Q &= Q^0(r_c) + Q^{S^2}.
\end{align}
\end{subequations}
Note that we use the gauge choice from Ref.~\cite{Balmelli:2015zsa}, i.e., there are no corrections of the form $\bm{S}\cdot\bm{S}$ in the potential $B_p$, which simplifies the Hamiltonian for aligned spins and circular orbits.

The 4PN corrections to the nonspinning effective Hamiltonian were obtained in Ref.~\cite{Damour:2015isa}.
Since we factor the PN corrections in $A^0(r_c)$, we choose it such that the PN expansion of the $A$ potential agrees, in the nonspinning limit, with the results of Ref.~\cite{Damour:2015isa}.
Writing the PN corrections using scaled variables \eqref{reducedVar} to simplify notation, we obtain 
\begin{subequations}
\label{PMcorr}
\begin{align}
\label{AMr}
A^0(r_c) &= 1 + \nu \bigg[\frac{2}{\hat{r}_c^3} + \left(\frac{106}{3} - \frac{41}{32} \pi^2\right) \frac{1}{\hat{r}_c^4}
+ \left(\frac{1}{20} + \frac{41}{32} \pi ^2 \nu - \frac{221}{6}\nu + \frac{963}{512}\pi^2 + \frac{128}{5}\gamma_E + \frac{256}{5}\ln 2 + \frac{64}{5} \ln \frac{1}{\hat{r}_c}\right) \frac{1}{\hat{r}_c^5}\bigg] 
\\
D^0(r_c) &= 1 + 6 \nu  \frac{1}{\hat{r}_c^2} + \left(52 \nu -6 \nu ^2\right) \frac{1}{\hat{r}_c^3} \nonumber\\
&\quad+ \left[\left(\frac{123}{16}\pi ^2 - 260\right) \nu ^2 + \nu  \left(-\frac{23761}{1536} \pi^2- \frac{533}{45}+\frac{1184 }{15} \gamma_E - \frac{6496}{15}\ln 2+\frac{2916}{5}\ln 3\right)+\frac{592}{15} \nu  \ln \frac{1}{\hat{r}_c} \right]\frac{1}{\hat{r}_c^4},
\label{DMr}  \\
Q^0(r_c) &= \left[ 2(4-3\nu)\nu \frac{1}{\hat{r}_c^2} + 
\left(\left(-\frac{5308}{15} + \frac{496256}{45} \ln 2 - \frac{33048}{5} \ln 3 \right) \nu 
- 83 \nu^2 + 10 \nu^3\right)\frac{1}{\hat{r}_c^3}
\right] \hat{p}_r^4 \nonumber\\
&\quad + \left[
\left(-\frac{827}{3} - \frac{2358912}{25} \ln 2 + \frac{1399437}{50} \ln 3 + \frac{390625}{18} \ln 5\right) \nu
- \frac{27}{5}\nu^2 + 6 \nu^3
\right] \frac{\hat{p}_r^6}{\hat{r}_c^2} \,,
\end{align}
\end{subequations}
\end{widetext}
where the corrections are expressed in terms of the centrifugal radius $r_c$, with $\hat r_c=r_c/M$, and $\hat p_r=p_r/\mu$.
Note that here and in the SEOB models discussed below, we are using Taylor-expanded and not resummed versions of these potentials---we want to compare the different ansätze of the Hamiltonians irrespective of possible resummations for the potentials (see also footnote \ref{nopade}).

Spin-squared contributions, up to NNLO, are added to the Hamiltonian using the following ansatz
\begin{subequations}
\label{S2ansatz}
\begin{align}
A^{SS} &= \frac{c_{n}}{\hat{r}_c^3}  \bm{\chi}_i \cdot \bm{\chi}_j  + \frac{c_{n}}{\hat{r}_c^4}  \bm{\chi}_i \cdot \bm{\chi}_j 
+ \frac{c_{n}}{\hat{r}_c^5}  \bm{\chi}_i \cdot \bm{\chi}_j, 
\\
A^{nS} &= \frac{c_{n}}{\hat{r}_c^3}  (\bm{n}\cdot\bm{\chi}_i)(\bm{n}\cdot\bm{\chi}_j) + \frac{c_{n}}{\hat{r}_c^4}  (\bm{n}\cdot\bm{\chi}_i)(\bm{n}\cdot\bm{\chi}_j) \nonumber\\
&\quad + \frac{c_{n}}{\hat{r}_c^5}  (\bm{n}\cdot\bm{\chi}_i)(\bm{n}\cdot\bm{\chi}_j),
\\
B_p^{nS} &= \frac{c_{n}}{\hat{r}_c^3}  (\bm{n}\cdot\bm{\chi}_i)(\bm{n}\cdot\bm{\chi}_j)
+ \frac{c_{n}}{\hat{r}_c^4}  (\bm{n}\cdot\bm{\chi}_i)(\bm{n}\cdot\bm{\chi}_j),
\\
B_{np}^{nS} &= \frac{c_{n}}{\hat{r}_c^3}  (\bm{n}\cdot\bm{\chi}_i)(\bm{n}\cdot\bm{\chi}_j)
+ \frac{c_{n}}{\hat{r}_c^4}  (\bm{n}\cdot\bm{\chi}_i)(\bm{n}\cdot\bm{\chi}_j),
\\
B_{np}^{SS} &= \frac{c_{n}}{\hat{r}_c^3}  \bm{\chi}_i \cdot \bm{\chi}_j 
+ \frac{c_{n}}{\hat{r}_c^4}  \bm{\chi}_i \cdot \bm{\chi}_j, \\
Q^{S^2} &= \frac{\hat{p}_r^4}{\hat{r}_c^3} \left[c_n \bm{\chi}_i \cdot \bm{\chi}_j  + c_n (\bm{n}\cdot\bm{\chi}_i)(\bm{n}\cdot\bm{\chi}_j)  \right] \nonumber\\
&\quad + \frac{\hat{p}_r^3}{\hat{r}_c^3} c_n (\bm{p}\cdot\bm{\chi}_i)(\bm{n}\cdot\bm{\chi}_j),
\end{align}
\end{subequations}
where we followed Ref.~\cite{Balmelli:2015zsa} in expressing the corrections in terms of $r_c$.
We employ notation such that, e.g., ${c_n \bm{\chi}_i \cdot \bm{\chi}_j} \equiv  c_n \bm{\chi}_1^2  + c_n \bm{\chi}_1 \cdot \bm{\chi}_2 + c_n \bm{\chi}_2^2 $.
Each $c_n$ stands for an independent undetermined coefficient in our ansatz, i.e., we use the same symbol $c_n$ for all coefficients to simplify notation. The full expressions after matching to PN results are provided in Appendix~\ref{app:coeffs}.
Note that we added LO S$^2$ corrections to the $A$-potential above (which vanish for black holes) to account for the multipole constants of neutron stars. 

The NLO S$^2$ contributions were included in the effective Hamiltonian in Ref.~\cite{Balmelli:2015zsa}, however the authors missed a contribution in matching the EOB Hamiltonian to PN results, namely from the LO S$^2$ generating function applied to the LO SO Hamiltonian, i.e., from the Poisson bracket $\lbrace \mathcal{G}^\text{LO}_{S^2}, H^\text{LO}_S \rbrace$.
In Appendix~\ref{app:BDcorrect}, we write the matching results for NLO S$^2$, using the notation of Ref.~\cite{Balmelli:2015zsa}, after taking into account the missing Poisson bracket.

The leading-order quartic-in-spin terms $A^{S^4}$ are zero for black holes, since the Kerr Hamiltonian, with the mapping $\vct{a} = \vct{a}_1 + \vct{a}_2$, automatically reproduces them, but they are nonzero for other types of compact objects.
We take the most generic expression for the S$^4$ corrections
\begin{align}
\label{S4ansatz}
A^{S^4} &= \frac{1}{\hat{r}_c^5} \bigg[
c_n (\bm{\chi}_i\cdot\bm{\chi}_j) (\bm{\chi}_k\cdot\bm{\chi}_l) \nonumber\\
&\quad\qquad + c_n (\bm{\chi}_i\cdot\bm{\chi}_j) (\bm{n}\cdot\bm{\chi}_k) (\bm{n}\cdot\bm{\chi}_l) \nonumber\\
&\quad\qquad
+ c_n (\bm{n}\cdot\bm{\chi}_i) (\bm{n}\cdot\bm{\chi}_j) (\bm{n}\cdot\bm{\chi}_k) (\bm{n}\cdot\bm{\chi}_l)
\bigg],
\end{align}
where a summation over the spins of the two bodies is implied, and terms symmetric under the exchange of the two bodies' labels are only included once.

The spin-orbit and spin-cubed PN corrections are added to the odd-in-spin part of the Kerr Hamiltonian $H_\text{odd}^\text{Kerr}$.
For the SO part we use the ansatz in Refs.~\cite{Balmelli:2015zsa,Damour:2014sva}, and we add to it S$^3$ corrections,
\begin{subequations}
\begin{align}
\label{HSOBD}
\hat H_\text{odd}^\text{eff} &= \frac{G_S}{\hat{r} \hat{r}_c^2 \left(1 + \Delta \frac{(\bm{n}\cdot\bm{a})^2}{{r}^2 {r}_c^2}\right)} \left(X_1^2 \hat{\bm{L}}\cdot \bm{\chi}_1 + X_2^2 \hat{\bm{L}}\cdot\bm{\chi}_2\right) \nonumber\\
&\quad 
+ \frac{G_{S^*}}{\hat{r}_c^3} \nu \left(\hat{\bm{L}}\cdot \bm{\chi}_1 + \hat{\bm{L}}\cdot\bm{\chi}_2\right) \nonumber\\
&\quad
+ \frac{G_{S^3}}{\hat{r}_c^4} \hat{\bm{L}}\cdot\bm{\chi}_1 + \frac{\tilde{G}_{S^3}}{\hat{r}_c^4}\hat{\bm{L}}\cdot\bm{\chi}_2 \,,
\end{align}
\end{subequations}
where
\begin{subequations} 
\begin{align}
\label{GSO}
G_S &= 2 \left[1 + \frac{c_n}{\hat{r}_c} + c_n \hat{p}_r^2 + \frac{c_n}{\hat{r}_c^2} + c_n \frac{\hat{p}_r^2}{\hat{r}_c} + c_n \hat{p}_r^4 \right]^{-1}, \nonumber\\
G_{S^*} &= \frac{3}{2} \left[1 + \frac{c_n}{\hat{r}_c} + c_n \hat{p}_r^2 + \frac{c_n}{\hat{r}_c^2} + c_n \frac{\hat{p}_r^2}{\hat{r}_c} + c_n \hat{p}_r^4 \right]^{-1} \\
%%%%%%%%
G_{S^3} &= \frac{1}{\hat{r}_c} \Big[ c_n \bm{\chi}_1\cdot\bm{\chi}_1 + c_n \bm{\chi}_2\cdot\bm{\chi}_2 + c_n \bm{\chi}_1\cdot\bm{\chi}_2  \nonumber\\
&\quad\qquad + c_n (\bm{n}\cdot\bm{\chi}_1)^2 
+ c_n (\bm{n}\cdot\bm{\chi}_2)^2 \nonumber\\
&\quad\qquad+ c_n (\bm{n}\cdot\bm{\chi}_1) (\bm{n}\cdot\bm{\chi}_2)\Big] \nonumber\\
&\quad +\hat{p}_r^2 \Big[c_n (\bm{n}\cdot\bm{\chi}_1)^2  + c_n (\bm{n}\cdot\bm{\chi}_2)^2  \nonumber\\
&\quad\qquad+ c_n (\bm{n}\cdot\bm{\chi}_1) (\bm{n}\cdot\bm{\chi}_2) \Big]  \nonumber\\
&\quad +\frac{\hat{L}^2}{\hat{r}^2} \Big[c_n (\bm{n}\cdot\bm{\chi}_1)^2  + c_n (\bm{n}\cdot\bm{\chi}_2)^2 \nonumber\\
&\quad\qquad + c_n (\bm{n}\cdot\bm{\chi}_1) (\bm{n}\cdot\bm{\chi}_2) \Big], \nonumber\\
\tilde{G}_{S^3} &= G_{S^3} \quad\text{with}\quad 1 \leftrightarrow 2.
\label{GS3}
\end{align}
\end{subequations}
Note that an inverse-Taylor resummation is used for $G_S$ and $G_{S^\ast}$, which improves the description of the binary dynamics for aligned spins \cite{Damour:2014sva}.
In the spin-cubed corrections $G_{S^3}$ and $\tilde{G}_{S^3}$, a gauge freedom exists which we chose such that terms of the form $p_r^2 \bm{\chi}_i\cdot\bm{\chi}_j$ or $L^2 \bm{\chi}_i\cdot\bm{\chi}_j$ are not included.
Explicit results after matching at 4PN can be found in Appendix~\ref{app:coeffsBD}.

\subsection{A simplified effective-one-body Hamiltonian with test-mass limit: {\simp}}

Since it is important to have fast and simple EOB waveform models, in this section, we consider a simplified version of the {\BD} Hamiltonian that uses $r$ instead of $r_c$ for the PN corrections.
In order to assess the effect of this simplification, we also avoid resummations that are not motivated by the structure of the interactions, i.e., we factorize spin corrections to the Kerr potentials and do not use an inverse-Taylor resummation for the spin-orbit part.

The potentials of the effective Hamiltonian are simply taken to be
\begin{subequations}
\label{S2S4simp}
\begin{align}
A &= A^\text{Kerr} \left(A^0 +  A^{SS} + A^{nS} + A^{S^4}\right), \\
B_p &= B_p^\text{Kerr} \left(1 + B_p^{nS}\right), \\
B_{np} &= \frac{ \left(1 - \frac{2}{\hat{r}} + \frac{\hat{a}^2}{\hat{r}^2}\right) \left(A^0 D^0 +  B_{np}^{SS} + B_{np}^{nS} \right)  - 1}{1 + (\bm{n}\cdot\hat{\bm{a}})^2/\hat{r}^2}, \\
B_{npa} &= B_{npa}^\text{Kerr}, \\
Q &= Q^0 + Q^{S^2},
\end{align}
\end{subequations}
where the zero-spin corrections $A^0(r),~D^0(r)$ and $Q^0(r)$ are given by Eq.~\eqref{PMcorr} but in terms of $r$ instead of $r_c$. The ans{\"a}tze for the S$^2$ and S$^4$ corrections are given by the corresponding expressions from the previous section, i.e., Eqs.~\eqref{S2ansatz} and \eqref{S4ansatz}, but using $r$ instead of $r_c$ (and with different coefficients $c_n$).
For $H_\text{odd}^\text{Kerr}$, we modify the odd-in-$a$ part of the Kerr Hamiltonian by the SO and S$^3$ PN corrections, that is
\begin{subequations}
\begin{align}
\label{HSOsimp}
\hat H_\text{odd}^\text{eff} &= \frac{1}{\hat{r} \hat{r}_c^2 \left(1 + \Delta \frac{(\bm{n}\cdot\bm{a})^2}{{r}^2 {r}_c^2}\right)} \Big[G_S \left(X_1^2 \hat{\bm{L}}\cdot \bm{\chi}_1 + X_2^2 \hat{\bm{L}}\cdot\bm{\chi}_2\right) \nonumber\\
&\qquad\quad +  G_{S^*}\nu \left(\hat{\bm{L}}\cdot \bm{\chi}_1 +\hat{\bm{L}}\cdot\bm{\chi}_2\right)  \nonumber\\
&\qquad\quad  + \frac{G_{S^3}}{\hat r} \hat{\bm{L}}\cdot\bm{\chi}_1 + \frac{\tilde{G}_{S^3}}{\hat r} \hat{\bm{L}}\cdot\bm{\chi}_2 \Big],
\end{align}
\end{subequations}
with the ans{\"a}tze for the coefficients given in Eqs.~\eqref{GSO} and \eqref{GS3}, but again written with $r$ instead of $r_c$ (and different $c_n$).
Explicit results after matching at 4PN can be found in Appendix~\ref{app:coeffsSimple}.

\subsection{Aligned effective-one-body Hamiltonian with test-mass limit and centrifugal radius: {\DN}}

In this section, we consider the aligned-spin EOB Hamiltonian proposed by Damour and Nagar in Ref.~\cite{Damour:2014sva} and extended in Refs.~\cite{Akcay:2018yyh,Nagar:2018plt,Nagar:2018zoe}, which we denote {\DN}. That Hamiltonian is similar to the aligned-spin limit of the {\BD} Hamiltonian from above, except that the even-in-spin PN corrections are added to the centrifugal radius.

The even-in-spin effective Hamiltonian is given by
\begin{align}
\hat H^\text{eff}_\text{even} &= \sqrt{A \left(1 + \frac{\hat L^2}{\hat r_c^2} + \frac{\hat p_r^2}{B} + Q^0\right)} .
\end{align}
The EOB potentials $A$ and $B$ are given in Ref.~\cite{Damour:2014sva}, but we do not use Pad\'{e} resummation and we modify how the zero-spin PN corrections are added such that they agree with the other Hamiltonians in this paper, that is 
\begin{subequations}
\begin{align}
A &= \left(1- \frac{2}{\hat{r}_c}\right) \frac{1 + \frac{2}{\hat{r}_c}}{1 + \frac{2}{\hat{r}_c}} A^0(r_c), \\
B &= \frac{r^2}{r_c^2} \frac{1}{A\, D^0(r_c)},
\end{align}
\end{subequations}
where $A^0$, $D^0$ and $Q^0$ are given by Eq.~\eqref{PMcorr}.
Note that Refs.~\cite{Nagar:2018plt,Nagar:2018zoe,Rettegno:2019tzh} use $Q \equiv 2\nu(4-3\nu) p_r^4/r_c^2$ instead of $Q^0$, and use $p_{r_\ast} \equiv p_r \sqrt{A/B}$ instead of $p_r$.

The spin-squared and spin-quartic corrections are added to the centrifugal radius, which is here defined by
\begin{equation}
\label{S2S4DN}
\hat{r}_c^2 = \hat{r}^2 + \hat{a}_Q^2 \left(1 + \frac{2}{\hat{r}}\right) + \frac{ \delta a^2_\text{NLO}}{\hat{r}} + \frac{\delta a^2_\text{NNLO}}{\hat{r}^2} + \frac{\delta a^4_\text{LO}}{\hat{r}^2},
\end{equation}
and where $\hat{a}_Q$ depends on the compact object's multipolar constants 
\begin{equation}
\hat{a}_Q^2 \equiv \hat a^2 + \tilde C_{1(ES^2)} \hat{a}_1^2 + \tilde C_{2(ES^2)} \hat{a}_2^2 ,
\end{equation}
where $\hat{a}_i = a_i / M$, $\hat{a} = |\vct{a}| / M$, and recalling Eq.~\eqref{BDspinmap}.

The spin-orbit part was obtained in Ref.~\cite{Damour:2014sva} with NNNLO $\nu$-independent spinning-test-body contributions and $\nu$-dependent contributions calibrated to NR. We do not include those higher-order corrections here, but we follow Ref.~\cite{Damour:2014sva} in using an inverse-Taylor resummation/calibration of the coefficients $G_S$ and $G_{S^*}$ in
\begin{align}
\label{SOS3DN}
\hat H_\text{odd}^\text{eff} &= \frac{G_S}{\hat{r} \hat{r}_c^2} \left(X_1^2 \hat{L} \chi_1 + X_2^2 \hat{L}\chi_2\right) 
+ \frac{G_{S^*}}{\hat{r}_c^3} \nu \left(\hat{L}\chi_1 + \hat{L}\chi_2\right) \nonumber\\
&\quad + \frac{G_{S^3}}{\hat{r}_c^4} \hat{L}\,\chi_1 + \frac{\tilde{G}_{S^3}}{\hat{r}_c^4}\hat{L}\,\chi_2 \,,
\end{align}
see Eqs.~\eqref{inversetaylorG}, and where $G_{S^3}$ and $\tilde{G}_{S^3}$ for aligned spins take the simple form
\begin{align}
G_{S^3} &= \frac{1}{\hat{r}_c} \left(c_n \chi_1^2 + c_n \chi_1\chi_2\right), \nonumber\\
\tilde{G}_{S^3} &= \frac{1}{\hat{r}_c} \left(c_n \chi_2^2 + c_n \chi_1\chi_2\right).
\end{align}
Including spin-cubic contributions was discussed in Appendix A of Ref.~\cite{Nagar:2018plt}, which we implement here so that the effective Hamiltonian includes all PN information at the same order as the other Hamiltonians considered in this paper.
Explicit results after matching at 4PN can be found in Appendix~\ref{app:coeffsDN}.

\section{Effective-one-body Hamiltonian with test-spin limit: {\BB}}
\label{sec:BB}

The SEOB Hamiltonian proposed in Refs.~\cite{Barausse:2009xi,Barausse:2011ys} is based on the Hamiltonian of a \emph{spinning} test body in the background of a Kerr black hole, which we here denote by {\BB} (see Table~\ref{tab:SEOBdefs}). 
In this section, we extend that Hamiltonian to 4PN order; compared to previous results, we add NLO S$^2$, NNLO S$^2$ LO S$^3$, and LO S$^4$ PN corrections, for generic compact objects and spin orientations.

The {\BB} Hamiltonian, as expressed in Ref.~\cite{Hinderer:2013uwa}, is given by
\begin{subequations}
\label{eq:Heffform}
\begin{align}
\hat{H}^\text{eff} &= \hat{H}^\text{eff}_\text{even} + \hat{H}^\text{eff}_\text{odd} + \hat{H}^\text{eff}_{S_*}, \label{eq:Hgeneric}\\
\hat{H}^\text{eff}_\text{even}&=\alpha \sqrt{q + Q^0} ,\\ 
\hat{H}^\text{eff}_\text{odd}&=\beta \hat{p}_\phi , \\
\hat{H}^\text{eff}_{S_*}&=\left[\hat{\bm{F}}_t+\left(\beta+\frac{\alpha \gamma^{\phi\phi} \hat{p}_\phi}{\sqrt{q}}\right)\hat{\bm{F}}_\phi\right]\cdot \hat{\bm{S}}_*\nonumber\\
&\quad +\frac{\alpha
}{\sqrt{q}}\left(\gamma^{rr}\hat{p}_r \hat{\bm{F}}_r+
\gamma^{\theta\theta}\hat{p}_\theta \hat{\bm{F}}_\theta\right)\cdot \hat{\bm{S}}_* \nonumber\\
&\quad
+\frac{1}{2 \hat{r}^3} \left[3 (\hat{\bm{S}}_* \cdot \bm{n})^2-\hat{\bm{S}}_*\cdot \hat{\bm{S}}_*\right],
\end{align}  
\end{subequations}
where 
\begin{equation}
q = 1+\gamma^{\phi\phi} \hat{p}_\phi^2+\gamma^{rr}\hat{p}_r^2+\gamma^{\theta \theta}\hat{p}_\theta^2,
\end{equation}
and $\hat{\bm{S}}_* \equiv \bm{S}_*/M\mu$ is a rescaling of the spin of the test body.
The spins are mapped according to
\begin{subequations}
\begin{align}
\bm{S}_*&={\bm{ \sigma}}_* \left[1+\nu f_{*}(r, \mathbf{p}) \right]+ \nu g_{*}(r, \mathbf{p}) {\bm{ \sigma}}, \\
{\bm{ \sigma}} &= \bm{S}_1+\bm{S}_2 \ , \\
{\bm \sigma}_* &= \frac{m_2}{m_1}\bm{S}_1+\frac{m_1}{m_2}\bm{S}_2,
\end{align}
\end{subequations}
where the functions $f_*$ and $g_*$ are given by Eqs.~(50)-(52) of Ref.~\cite{Barausse:2011ys}, and that ${\bm{ \sigma}} = M \vct{a}$ is the spin of the background Kerr metric;
it does not hold $\vct{a} = \vct{a}_1 +  \vct{a}_2$ as for the models discussed above.
The spin maps are analogous to the mapping of the masses $M$, $\mu$ according to Eq.~\eqref{massmap}, with the difference that the spin maps relate dynamical variables.
The deformed metric is obtained by substituting $\Delta$, $\Sigma$, and $\Lambda$ in the Kerr metric by
\begin{subequations}
\begin{align}
\Delta_t&=\hat{r}^2 A(r)+\hat{\sigma}^2, \\
\Delta_r &=\Delta_t D^0(r), \\
\hat{\Sigma}&=\hat{r}^2+\hat{\sigma}^2 \cos^2\theta, \\
\Lambda_t&=(\hat{r}^2+\hat{\sigma}^2)^2-\hat{\sigma}^2 \Delta_t \sin^2\theta,
\end{align}
\end{subequations}
as in
\begin{subequations}
\begin{gather}
\alpha = \frac{\sqrt{\Delta_t \hat{\Sigma}}}{\sqrt{\Lambda_t}}, \qquad \beta=\frac{2\hat{\sigma} \hat{r}}{\Lambda_t},\\
 \gamma^{\phi\phi} = \frac{\hat{\Sigma}}{\Lambda_t\sin^2\theta},  \quad
 \gamma^{rr}=\frac{\Delta_r}{\hat{\Sigma}}, \quad \gamma^{\theta\theta}=\frac{1}{\hat{\Sigma}}.
\end{gather}
\end{subequations}
The potential $D^0(r)$ is given by Eq.~\eqref{DMr}, and the potential $A$ is given by 
\begin{equation}
A = \hat{a}^2\left(\frac{1}{\hat{r}} - \frac{1}{\hat{r}_{H,+}}\right)\left(\frac{1}{\hat{r}} - \frac{1}{\hat{r}_{H,-}}\right) A^0(r) - \frac{\hat{a}^2}{\hat{r}^2},
\end{equation}
where $A^0$ is given by Eq.~\eqref{AMr}, and $\hat{r}_{H,\pm}$ are the scaled inner and outer radii of a Kerr black hole, i.e.,
\begin{equation}
\hat{r}_{H,\pm} = 1 \pm \sqrt{1 - \hat{a}^2} \,.
\end{equation}

Finally, the vectors $\hat{\bm{F}}_t$, $\hat{\bm{F}}_r$, $\hat{\bm{F}}_\theta$, and $\hat{\bm{F}}_\phi$ describe the fictitious force acting on the test-body spin $\hat{\bm{S}}_*$ (frame dragging) in the deformed Kerr metric.
They are given by Eq.~(6) in Ref.~\cite{Hinderer:2013uwa}, which we rewrite here for convenience,
\begin{widetext}
\begin{subequations}
\begin{align}
\hat{\bm{F}}_\phi &=\cos\theta  \ \bm{\hat{ n}}+ \bm{\hat{ v}},\\
%%%%%%%
\hat{\bm{F}}_t&= \bm{\hat{n}} \ \frac{\sqrt{\gamma^{\phi\phi}}\sqrt{ \gamma^{\theta\theta}}}{\sqrt{q}}\left[\frac{\hat{p}_\phi \alpha_{,\theta}(1+2\sqrt{q})}{(1+\sqrt{q})}-\alpha \hat{p}_\phi \cot \theta-\frac{(1-2\sqrt{q})\beta_{,\theta}}{2\gamma^{\phi\phi}}\right]\nonumber\\
&\quad +\bm{\hat{ v}} \ \frac{\csc\theta\sqrt{\gamma^{rr}}}{\sqrt{\gamma^{\phi\phi}}}\left[\frac{\gamma^{\phi\phi} \hat{p}_\phi \alpha_{,r}}{(1+\sqrt{q})}+\frac{(2\sqrt{q}-1)\beta_{,r}+\alpha \hat{p}_\phi \gamma^{\phi\phi}_{,r}}{2\sqrt{q}}\right], \; \; \; \\
%%%%%%
\hat{\bm{F}}_r &= -\bm{\hat{ n}} \ \frac{\sqrt{\gamma^{\theta\theta}}(\beta_{,\theta}\hat{p}_r+\beta_{,r}\hat{p}_\theta)}{2\alpha \sqrt{\gamma^{\phi\phi}}(1+\sqrt{q})}- \bm{\hat{ v}} \ \frac{\csc\theta\left(\beta_{,\theta} \gamma^{\theta \theta}\hat{p}_\theta+2 \hat{p}_r \gamma^{rr} \beta_{,r}\right)}{2\alpha \sqrt{\gamma^{\phi\phi}} \sqrt{\gamma^{rr}}(1+\sqrt{q})}- \bm{\hat{ \xi}} \ \frac{\csc\theta \sqrt {\gamma^{\theta \theta}}}{2\alpha \sqrt{\gamma^{rr}}}\left[\frac{2\sqrt{q}\alpha_{,\theta}+\hat{p}_\phi \beta_{,\theta}}{ (1+\sqrt{q})}+\frac{\alpha \gamma^{\theta\theta}_{,\theta}}{\gamma^{\theta\theta}}\right],\ \ \\
%%%%%%
\hat{\bm{F}}_\theta &= - \bm{\hat{ n}}  \ \frac{\sqrt{\gamma^{\theta \theta}}\beta_{,\theta}\hat{p}_\theta}{\alpha \sqrt{\gamma^{\phi\phi}}(1+\sqrt{q})}- \bm{\hat{ v}} \ \frac{\csc\theta\sqrt{\gamma^{rr}} \hat{p}_\theta \beta_{,r}}{2\alpha \sqrt{\gamma^{\phi\phi}}(1+\sqrt{q})}+\bm{\hat{ \xi}}  \csc\theta \ \bigg[1+\frac{\sqrt{\gamma^{rr}}}{2\alpha\sqrt{\gamma^{\theta \theta}}}\bigg(\frac{2\sqrt{q} \alpha_{,r}+\hat{p}_\phi \beta_{,r}}{ (1+\sqrt{q})}+\frac{\alpha \gamma^{\theta \theta}_{ \; ,r}}{\gamma^{\theta \theta}} \bigg)\bigg]. \; \; \; \; 
\end{align}
\end{subequations}
Here, the unit vectors \mbox{$(\bm{\hat{ n}},\bm{\hat{ \xi}},\bm{\hat{ v}})$} are defined by
$\hat{\bm n}=\frac{{\bm x}}{r}$, $\hat{\bm \xi}=\hat e_{\rm Z}^{\rm \sigma}\times \hat {\bm n}$, and $\hat {\bm v}=\hat{\bm n}\times \hat {\bm \xi}$,
where \mbox{$\hat{e}_{\rm Z}^{\rm \sigma}={\bm \sigma}/\sigma$} denotes the direction of the (deformed) 
Kerr spin.

For the purpose of extending the {\BB} Hamiltonian to 4PN in the spinning sector, we deviate from the original philosophy of Refs.~\cite{Barausse:2009xi,Barausse:2011ys} in that we do not modify the spin maps or deform the metric entering $H^\text{eff}_{S_*}$ with terms of quadratic and higher order in spin.
Instead, we only slightly modify the ansatz for the effective Hamiltonian (keeping $H^\text{eff}_{S_*}$ unchanged) as
\begin{subequations}
\label{HBBansatz}
\begin{align}
\hat{H}^\text{eff}_\text{even}&=\sqrt{\alpha^2 + A^{SS} + A^{nS} + A^{S^4}} \sqrt{q + B_p^{nS} \hat{\vct{p}}^2 + (B_{np}^{SS}+B_{np}^{nS}) \hat{p}_r^2 + Q^0 + Q^{S^2} + Q^{S^4}} \\
\hat{H}^\text{eff}_\text{odd}&=\beta \hat{p}_\phi + \frac{G_{S^3}}{\hat{r}^4} \hat{\bm{L}}\cdot\bm{\chi}_1 + \frac{\tilde{G}_{S^3}}{\hat{r}^4}\hat{\bm{L}}\cdot\bm{\chi}_2 ,
\label{HoddBB}
\end{align}  
\end{subequations}
where we introduced potentials into $H^\text{eff}_\text{even}$ at quadratic and higher order in spin following the structure of Eq.~\eqref{Heffansatz}.
These potentials $A^{SS},\, A^{nS},\, B_p^{nS},\, B_{np}^{SS},\,B_{np}^{nS},\, Q^{S^2}$ and $A^{S^4}$ are given by Eqs.~\eqref{S2ansatz} and \eqref{S4ansatz} but with $r$ instead of $r_c$ (and different $c_n$), and similarly for the spin-cubic corrections $G_{S^3}$ and $\tilde{G}_{S^3}$ from Eq.~\eqref{GS3}.
We take the function $Q^{S^4}$ to have the form
\begin{align}
Q^{S^4} &= \frac{\hat{p}_r^2}{\hat{r}^4} \bigg[
c_n (\bm{n}\cdot\bm{\chi}_1)^3(\bm{n}\cdot\bm{\chi}_2)
+ c_n (\bm{n}\cdot\bm{\chi}_1)^2 (\bm{\chi}_1\cdot\bm{\chi}_2)
+ c_n (\bm{\chi}_1\cdot\bm{\chi}_1)(\bm{n}\cdot\bm{\chi}_1)(\bm{n}\cdot\bm{\chi}_2)
+ c_n (\bm{n}\cdot\bm{\chi}_2)^3(\bm{n}\cdot\bm{\chi}_1) \nonumber\\
&\qquad\quad + c_n (\bm{n}\cdot\bm{\chi}_2)^2 (\bm{\chi}_1\cdot\bm{\chi}_2)
+ c_n  (\bm{\chi}_2\cdot\bm{\chi}_2)(\bm{n}\cdot\bm{\chi}_1)(\bm{n}\cdot\bm{\chi}_2)
+ c_n (\bm{n}\cdot\bm{\chi}_1)^2(\bm{n}\cdot\bm{\chi}_2)^2
+ c_n (\bm{n}\cdot\bm{\chi}_1)^2(\bm{\chi}_2\cdot\bm{\chi}_2) \nonumber\\
&\qquad\quad +c_n (\bm{\chi}_1\cdot\bm{\chi}_1) (\bm{\chi}_2\cdot\bm{\chi}_2)
+ c_n (\bm{\chi}_1\cdot\bm{\chi}_1) (\bm{n}\cdot\bm{\chi}_2)^2\bigg] \nonumber\\
&\quad +\bm{p}\cdot\bm{\chi}_1  \frac{\hat{p}_r}{\hat{r}^4}
\bigg[ c_n (\bm{\chi}_2\cdot\bm{\chi}_2)(\bm{n}\cdot\bm{\chi}_2)
+ c_n (\bm{n}\cdot\bm{\chi}_2)^3
+ c_n (\bm{n}\cdot\bm{\chi}_1)^2 (\bm{n}\cdot\bm{\chi}_2)
+ c_n (\bm{n}\cdot\bm{\chi}_1) (\bm{\chi}_1\cdot\bm{\chi}_2)
+ c_n (\bm{n}\cdot\bm{\chi}_1) (\bm{n}\cdot\bm{\chi}_2)^2 \nonumber\\ 
&\qquad\qquad\qquad + c_n (\bm{n}\cdot\bm{\chi}_1) (\bm{\chi}_2\cdot\bm{\chi}_2)
\bigg] \nonumber\\
&\quad +\bm{p}\cdot\bm{\chi}_2  \frac{\hat{p}_r}{\hat{r}^4}
\bigg[ c_n (\bm{\chi}_1\cdot\bm{\chi}_1)(\bm{n}\cdot\bm{\chi}_1)
+ c_n (\bm{n}\cdot\bm{\chi}_1)^3
+ c_n (\bm{n}\cdot\bm{\chi}_2)^2 (\bm{n}\cdot\bm{\chi}_1)
+ c_n (\bm{n}\cdot\bm{\chi}_2) (\bm{\chi}_1\cdot\bm{\chi}_2)
+ c_n (\bm{n}\cdot\bm{\chi}_2) (\bm{n}\cdot\bm{\chi}_1)^2 \nonumber\\ 
&\qquad\qquad\qquad + c_n (\bm{n}\cdot\bm{\chi}_2) (\bm{\chi}_1\cdot\bm{\chi}_1)
\bigg].
\end{align}
\end{widetext}
Out of 38 possible terms in the most general expression for $Q^{S^4}$, 16 terms were removed via a gauge choice. We started by removing the three terms that do not vanish for aligned spins, but the other 13 terms were chosen arbitrarily.
Explicit results after matching at 4PN can be found in Appendix~\ref{app:coeffsBB}.

\begin{figure*}
\includegraphics[width=0.8\linewidth]{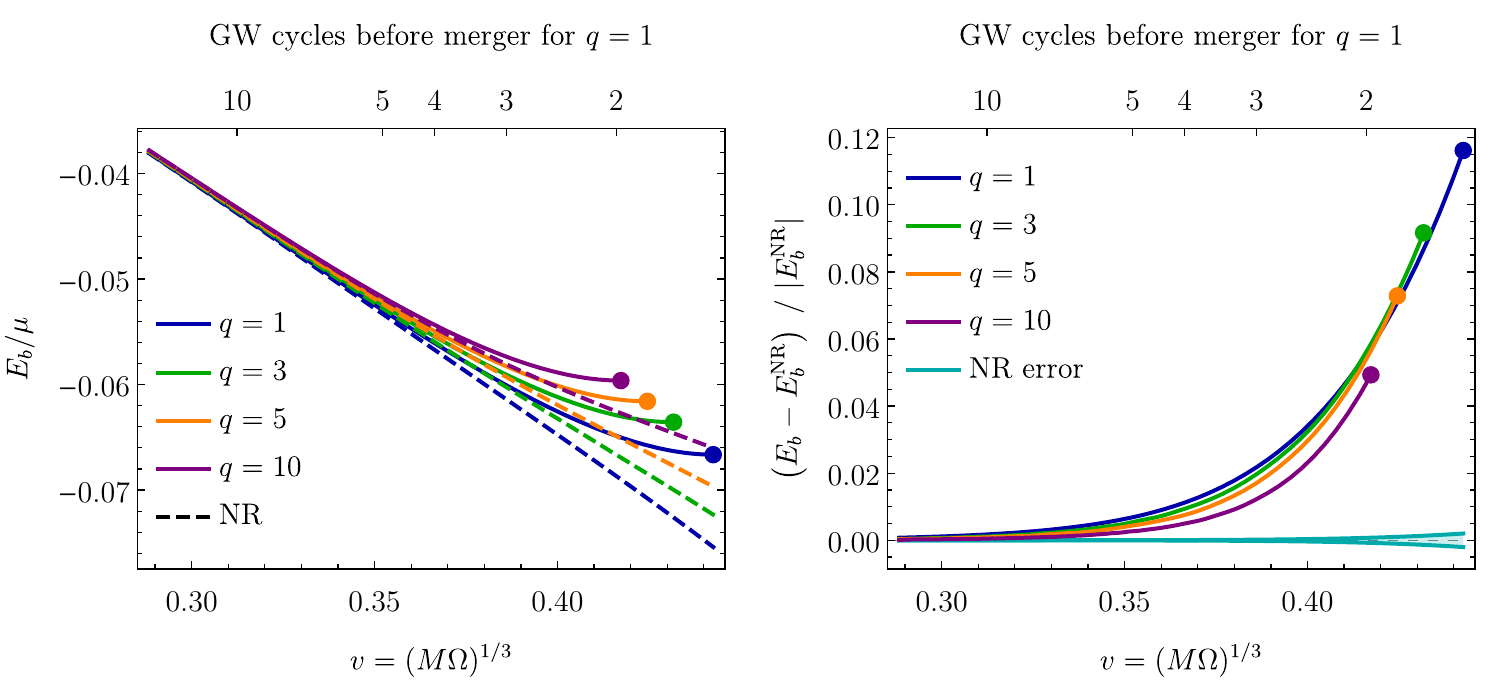}
\caption{\label{fig:nonspin} Binding energy (left panel)  and fractional binding energy (right panel) versus the ``velocity'' parameter $v$ for nonspinning binary--black-hole configurations with different mass ratios. The four SEOB Hamiltonians considered here are identical for zero spin. The relative NR error shown in the right panel is a conservative 1.1\% estimate. The initial value of $v$ (the left end of the plots' domain) here is determined by the beginning of the NR simulation with $q=10$; those with lower mass ratios have several cycles at lower frequencies not shown here. We stress that the SEOB Hamiltonians at 4PN order are not calibrated to NR simulations. }
\includegraphics[width=\linewidth]{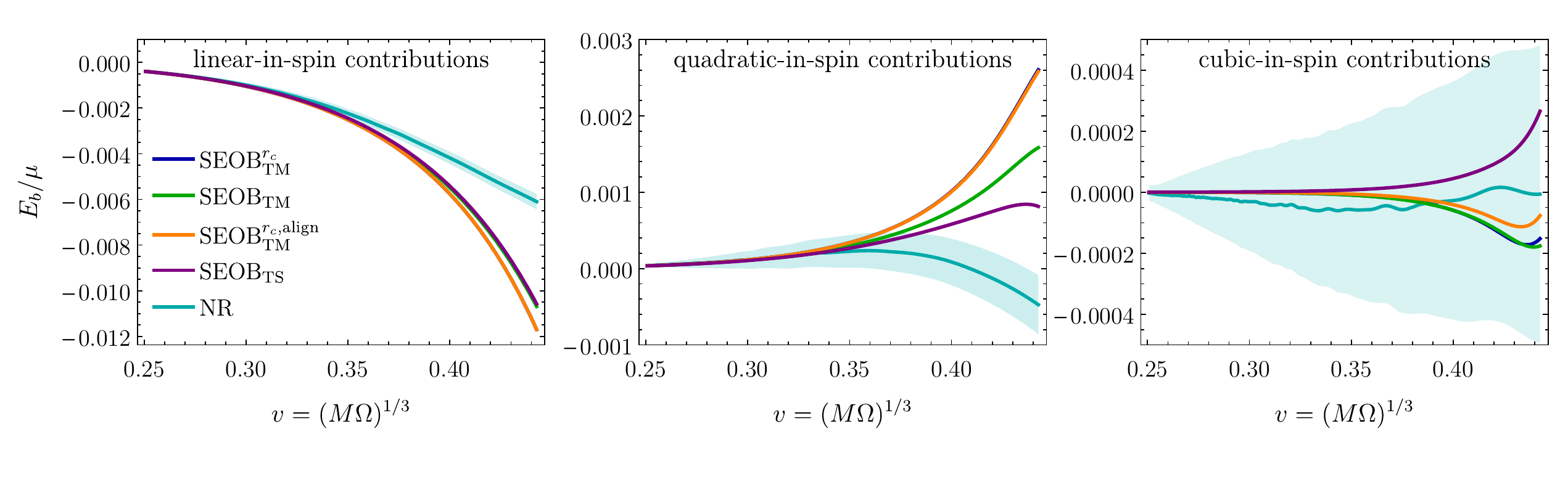}
\caption{\label{fig:spinCont} Binding energy versus the ``velocity'' parameter $v$ for the linear-in-spin (left panel), quadratic-in-spin (central panel) and cubic-in-spin (right panel) contributions of the four SEOB Hamiltonians. The NR error is indicated by the shaded regions. In the left panel, the blue and orange curves overlap since the {\BD} and {\DN} Hamiltonians are identical in the spin-orbit limit.}
\end{figure*}

\section{Comparison with numerical relativity}
\label{compare}

In this section, we compare the four SEOB Hamiltonians considered in this paper to NR simulations through the binding energy for circular orbits and aligned spins. The NR binding energy data we use here were extracted from the Simulating eXtreme Spacetimes (SXS) catalog \cite{SXS} in Ref.~\cite{Ossokine:2017dge}. Hereafter, in this section, we use the term ``aligned spins'' to mean spins parallel to, and in the same direction as, the orbital angular momentum $\bm{L}$, but we use the term ``antialigned spins'' to mean spins opposite to the direction of $\bm{L}$.

The binding energy is calculated by evaluating the EOB Hamiltonian for circular orbits ($p_r=0$) and solving numerically $\dot{p}_r=-\partial H_\text{EOB}/\partial r=0$ for the angular momentum $p_\phi$ at some radius.  
The orbital frequency $\Omega$ is obtained from
\begin{equation}
\Omega = \frac{\partial H_\text{EOB}}{\partial p_\phi} \,.
\end{equation}
We then calculate the binding energy and orbital frequency as $r$ goes from the beginning of the NR simulation to the innermost-stable circular orbit (ISCO) of the Hamiltonian, which marks the end of the inspiral phase of the binary coalescence and the beginning of the plunge.
The ISCO is calculated by setting both the first and second derivatives of the Hamiltonian with respect to $r$ to zero, i.e., 
$\partial H_\text{EOB}/\partial r = 0 = \partial^2 H_\text{EOB}/\partial r^2$.

It should be noted that the binding energy is extracted from NR simulations from an evolving binary, tracking the radiated energy in GWs. 
From the EOB Hamiltonians, however, we obtain the binding energy here by assuming exact circular orbits at different orbital separations, neglecting the orbital decay (radiation-reaction) due to the emitted GWs. The NR and EOB binding energies are thus not expected to agree exactly here during the last few orbits (see discussions in Ref.~\cite{Antonelli:2019ytb}).

\begin{figure*}
	\includegraphics[width=0.7\linewidth]{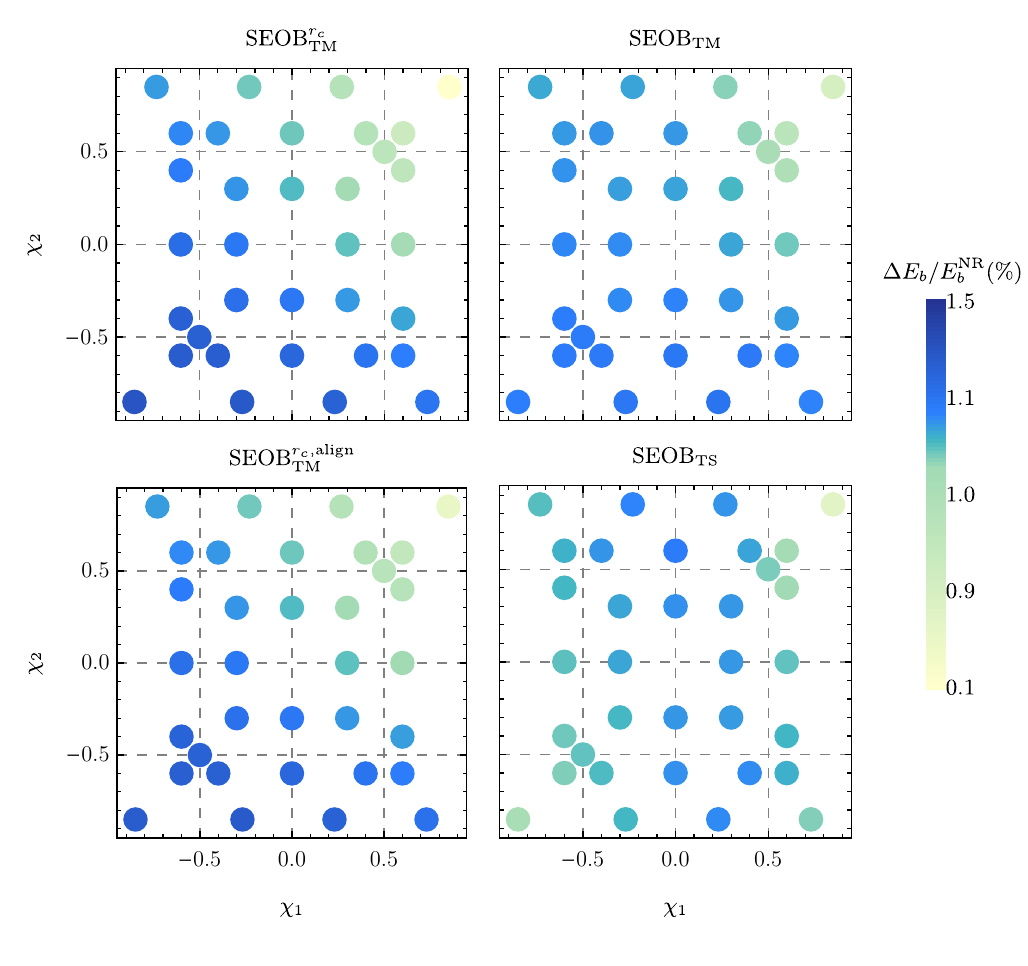}
	\caption{\label{fig:fracenergy} Fractional difference in the binding energy between NR and SEOB Hamiltonians for different spin configurations with mass ratio $q=3$ at 4 GW cycles before merger.}
\end{figure*}

In Fig.~\ref{fig:nonspin}, we plot the binding energy for nonspinning configurations with different mass ratios $q$ as a function of the ``velocity'' parameter $v\equiv (M\Omega)^{1/3}$, and we see that the binding energy increases with increasing mass ratio. The top axis of the figure indicates the number of GW cycles before merger, computed from the SXS waveform, for the case of $q=1$, which is close to the other values of $q$; for example, at 2 GW cycles before merger, $v=0.416$ for $q=1$ while $v=0.415$ for $q=10$.
Since all SEOB Hamiltonians considered here agree in the nonspinning limit by construction, this figure gives a rough estimate for the zero-spin contributions to the binding energy.
In all plots of this section the number of GW cycles from merger is always computed from the SXS waveforms, and the merger is 
defined as the peak of the (2,2) gravitational mode.

The different spin contributions to the binding energy are depicted in Fig.~\ref{fig:spinCont}.
They can be extracted by combining results for various spin combinations as (see Refs.~\cite{Dietrich:2016lyp, Ossokine:2017dge}) 
\begin{subequations}
\begin{align}
E_\text{SO} &= -\frac{1}{6}(-0.6, 0) + \frac{8}{3} (0.3, 0) - 2 (0, 0) - \frac{1}{2} (0.6, 0), \nonumber\\
E_{S^2} &= \frac{3}{2}(-0.6, 0) - 2 (0, 0) + \frac{3}{2} (0.6, 0)  - (0.6, -0.6), \nonumber\\
E_{S^3} &= -\frac{5}{6}(-0.6, 0) - \frac{8}{3} (0.3, 0) + 3 (0, 0) - \frac{1}{2} (0.6, 0) \nonumber\\
&\quad
 + \frac{1}{2} (0.6, -0.6) + \frac{1}{2} (0.6, 0.6),
\end{align}
\end{subequations}
where the numbers in brackets refer to the values of the dimensionless
spins of the two bodies ($\chi_1$, $\chi_2$).  The spin-squared
contributions to the binding energy $E_{S^2}$ refer to both $S_i^2$
and $S_1 S_2$ interactions. Similarly, spin-cubic contributions
$E_{S^3}$ refer to both $S_i^3$ and $S_i^2S_j$.  We see that the
spin-orbit contribution is about an order of magnitude larger than the
spin-squared contribution, which in turn is an order of magnitude
larger than the spin-cubic contribution.  All SEOB Hamiltonians give
comparable results for the spin-orbit part, however for the
spin-squared contribution, the {\BB} and {\simp} Hamiltonians give
better agreement with NR than the other two Hamiltonians.  For the
cubic-in-spin contributions, the NR error is larger than the EOB
values for the binding energy, and hence we cannot conclude which
Hamiltonian is better in terms of S$^3$ contributions.

\begin{figure*}
	\includegraphics[width=\linewidth]{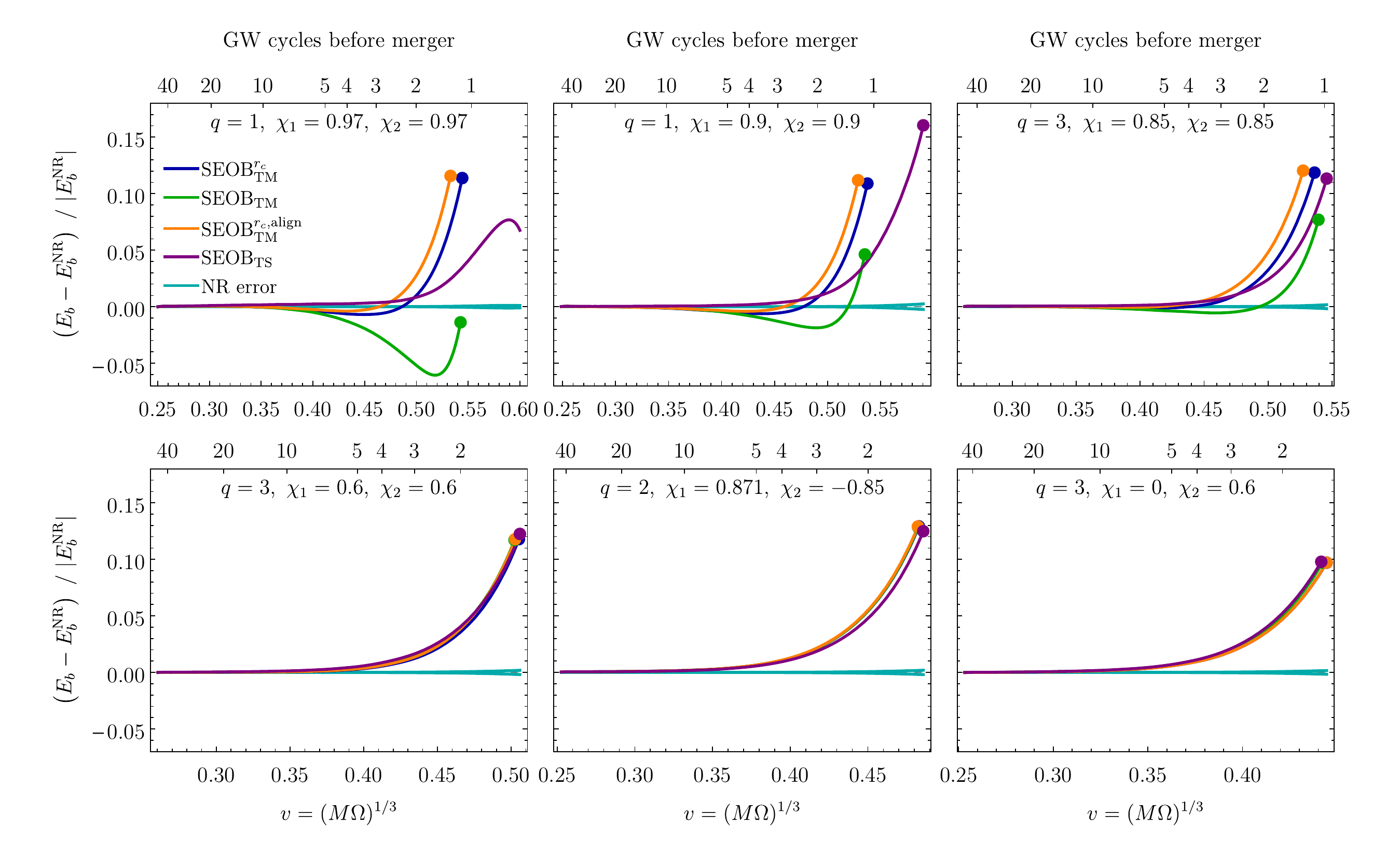}
	\caption{\label{fig:aligned} Binding energy comparison with NR for different aligned-spin configurations for the four SEOB 
Hamiltonians. Curves that end with a point indicate the location of the ISCO. For the NR error, we used 1.1\% relative error as a very conservative estimate.}
	
	\includegraphics[width=\linewidth]{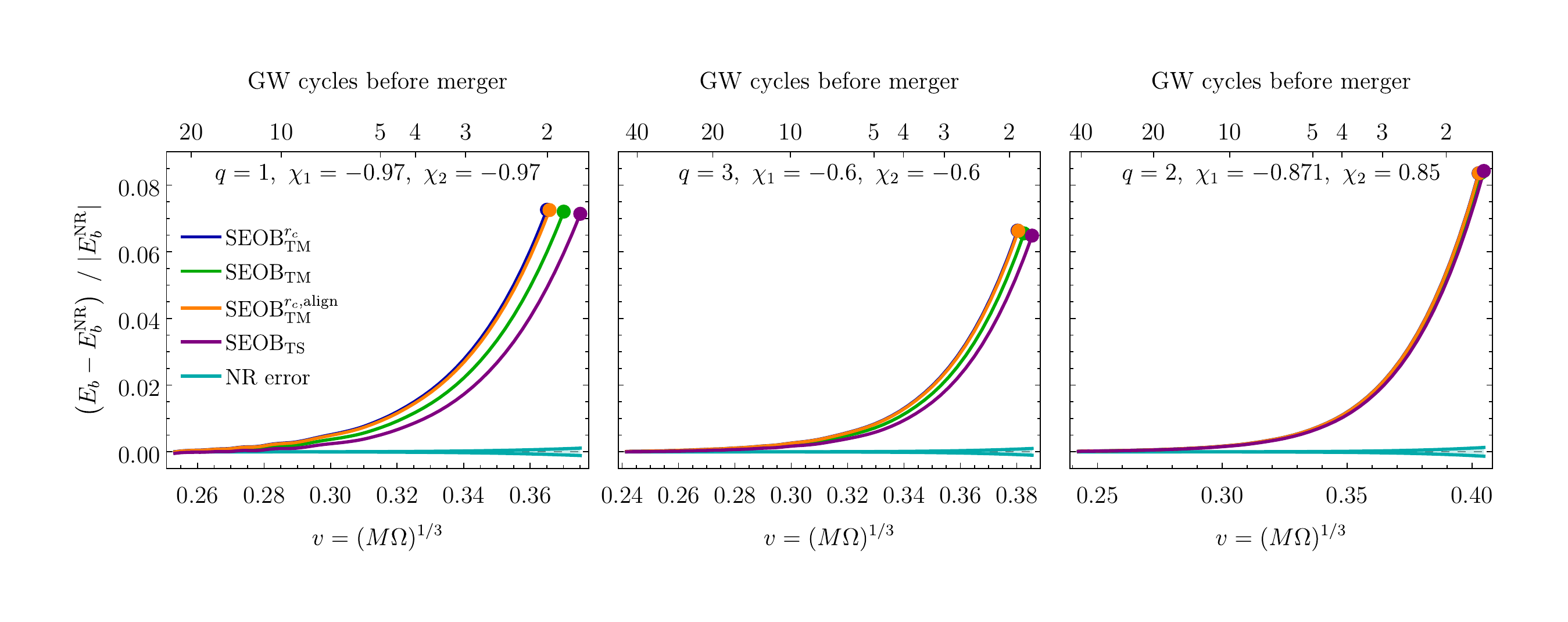}
	\caption{\label{fig:antialigned} As in Fig.~\ref{fig:aligned} but for configurations with antialigned spins.}
\end{figure*}

In Fig.~\ref{fig:fracenergy}, we compare the fractional energy difference $|E_b-E_b^\text{NR}|/E_b^\text{NR}$ at four GW cycles (i.e., two orbits) before merger for various spin configurations with mass ratio $q=3$.
We see that, for all configurations at that frequency, the relative difference with NR is around $1\%$.
For aligned spins, all Hamiltonians give comparable results, but the {\BB} Hamiltonian gives better agreement with NR for antialigned spins.

\begin{figure*}
	\includegraphics[width=0.7\linewidth]{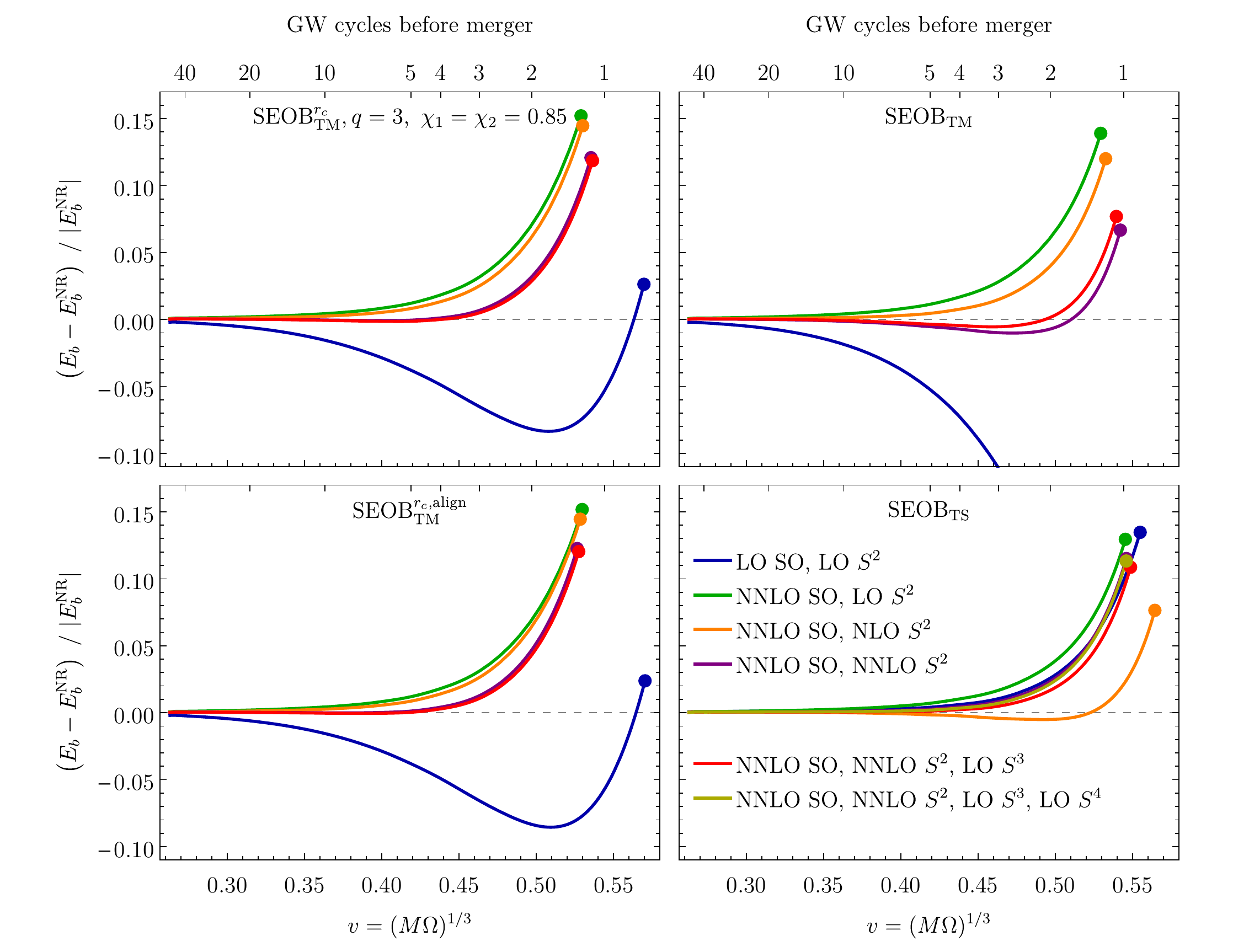}
	\caption{\label{fig:PNcomp} Comparing the effect of adding spin contributions to the binding energy at different PN orders for 
the four SEOB  Hamiltonians.}
\end{figure*}

We also compare the binding energy as a function of velocity for some configurations with aligned spins (Fig.~\ref{fig:aligned}) and antialigned spins (Fig.~\ref{fig:antialigned}).
The curves in these figures start at the beginning of the available NR simulations and end at the ISCO of the EOB Hamiltonians.
All effective Hamiltonians considered here have an ISCO for arbitrary spins, except that the {\BB} Hamiltonian does not have an ISCO for large aligned spins $\gtrsim 0.92$.
From the three panels at the top of Fig.~\ref{fig:aligned}, we see that for large aligned spins ($\gtrsim 0.8$), the {\BB} Hamiltonian shows slightly better agreement with NR than the other Hamiltonians. 
However, for smaller spins ($\lesssim 0.6$), all Hamiltonians give very similar results. This is also the case when the two spins are both large but in opposite directions.
For antialigned spins, the difference between the four Hamiltonians is smaller than in the aligned-spin case; the {\BB} Hamiltonian gives better agreement with NR than the other Hamiltonians, but the difference is small, even for spin magnitudes of $0.97$, and becomes negligible for smaller spins.

Finally, in Fig.~\ref{fig:PNcomp}, we compare the effect of adding spin PN orders to the effective Hamiltonian for a configuration with mass ratio $q=3$ and spins $\chi_1=\chi_2=0.85$.
For all Hamiltonians, adding higher spin orders improves agreement with NR, except for the {\BB} Hamiltonian where adding LO S$^3$ and LO S$^4$ gives slightly worse agreement.
We checked that using different spin configurations gives qualitatively similar behavior. 

Overall, beside the small differences pointed out above, all Hamiltonians perform reasonably well compared to NR simulations.
One should expect that the differences that accumulate during the last orbits can be compensated by a calibration of the Hamiltonians, applying also further resummations to the potentials, which we leave for future work.
This would be of particular interest for the simplified Hamiltonian, in order to prepare and evaluate it as a possible starting point for an EOB waveform model.

\section{Conclusions}
\label{conclude}

In this paper, we built spinning EOB Hamiltonians that include the
complete fourth post-Newtonian conservative dynamics for generic
(precessing) spins.  These Hamiltonians are also valid for generic
compact objects (e.g., black holes or neutron stars) since we included multipole
constants that parametrize the deformation of the compact object due
to its rotation.

In particular, we considered and extended four SEOB Hamiltonians:
(i) an extension of the SEOB Hamiltonian from
Ref.~\cite{Balmelli:2015zsa} by adding NNLO S$^2$ and LO S$^3$
contributions, in addition to adding the multipole constants; (ii) a
simplified version of that Hamiltonian that differs in how the spin
corrections are added to the Kerr metric, and that does not use the
concept of centrifugal radius; (iii) the aligned-spin Hamiltonian from
Refs.~\cite{Damour:2014sva,Nagar:2018plt,Nagar:2018zoe}, which
already includes complete 4PN information for generic compact objects,
but considered here for comparison with the other Hamiltonians;
(iv) an extension of the SEOB Hamiltonian from
Refs.~\cite{Barausse:2009xi,Barausse:2011ys}, which uses a test spin,
by adding NLO S$^2$, NNLO S$^2$, LO S$^3$, and LO S$^4$ contributions,
in addition to adding the multipole constants.  Since our goal in this
paper was to improve the description of spin effects in the EOB formalism, we
modified the zero-spin part of the above Hamiltonians such that they
are identical in that limit.  Furthermore, we did not include NR
calibration parameters or resummations of the PN corrections (e.g., with 
Pad\'{e} or $\log$ resummations of the zero-spin part) since they constitute an
implicit calibration to NR that improves the performance of EOB
Hamiltonians only in certain models.

We compared the four SEOB Hamiltonians considered here with NR
simulations by calculating the binding energy for circular orbits and
aligned spins. We found that all Hamiltonians show good agreement with
NR, and that the difference between the Hamiltonians is 
quite small up to moderate values of the spins and a handful 
number of GW cycles before merger. For large spins, the {\BB}
Hamiltonian performs better at large frequencies, but since all
Hamiltonians have an error of about $1\%$ compared to NR at about  four GW cycles
before merger, the simplest SEOB Hamiltonian \simp\, could be an excellent 
candidate for building an improved EOB waveform model with
precessing spins. The simplicity will allow one to have a 
fast-to-evolve set of equations of motion, and could help in calibrating 
the EOB waveforms built with \simp\, to NR simulations. However, more analyses,  
which include dissipative effects, and a careful study of how the GW frequency approaches 
merger, are needed to pin down the more suitable  SEOB Hamiltonian. Indeed, as several studies 
have shown~\cite{Taracchini:2012ig,Taracchini:2013rva,Pan:2013rra,Bohe:2016gbl,Babak:2016tgq,Cotesta:2018fcv}, 
to attach robustly the merger-ringdown waveform to the inspiral-plunge one in the EOB formalism, dynamical 
quantities, such as the orbital frequency, radial separation and momentum vectors, have 
to behave regularly around and beyond the EOB photon orbit.

We leave for future work to complete the SEOB Hamiltonians to a
gravitational waveform model, i.e., provide resummed expressions for
GW modes and associated radiation-reaction forces.
Once radiation-reaction forces are included in the model, it is
important to perform comparisons with NR for precessing spins and to
use those comparisons to study different resummation options and to
add calibration parameters in order to improve the accuracy of EOB
waveforms toward merger.

\section*{Acknowledgments}
We thank Sergei Ossokine for providing us with the numerical-relativity data for the binding energy used in this paper.

\appendix

\section{Completing the spinning effective-one-body Hamiltonian in Ref.~\cite{Balmelli:2015zsa} at NLO S$^2$}
\label{app:BDcorrect}

We found that Ref.~\cite{Balmelli:2015zsa} missed a contribution in the matching between the EOB Hamiltonian and PN results, namely from the LO spin-squared canonical transformation, generated by $G^\text{LO}_\text{SS}$, applied to the LO spin-orbit Hamiltonian $H^\text{LO}_\text{SO}$.
This contribution can be obtained using the Poisson bracket as $\{ G^\text{LO}_\text{SS}, H^\text{LO}_\text{SO} \}$.
This leads to SS contributions via the Poisson bracket of the spin vector $\{ S^i, S^j \} = \epsilon_{ijk} S^k$, which turns out to start at NLO in the SS sector.
Taking these additional contributions into account, we find that the coefficients in the EOB potentials $a^{\chi}_{ij}$, $a^{n\chi}_{ij}$, $b^{p,\chi}_{ij}$, $b^{p,n\chi}_{ij}$, $b^{np,\chi}_{ij}$, $b^{np,n\chi}_{ij}$, defined in Ref.~\cite{Balmelli:2015zsa} should read, assuming the gauge conditions $b^{p,\chi}_{ij} = 0$,
\begin{subequations}
\begin{align}
  a^{n\chi}_{11} &= 7 \nu X_1 + \frac{5}{4} \nu^2 , \\
  a^{n\chi}_{22} &= 7 \nu X_2 + \frac{5}{4} \nu^2 , \\
  a^{n\chi}_{12} &= a^{n\chi}_{21} = \frac{27}{8} \nu - \frac{9}{4} \nu^2 , \\
  b^{p,n\chi}_{11} &= 4 \nu X_1 - \frac{5}{2} \nu^2 , \\
  b^{p,n\chi}_{22} &= 4 \nu X_2 - \frac{5}{2} \nu^2 , \\
  b^{p,n\chi}_{12} &= b^{p,n\chi}_{21} = \frac{9}{8} \nu + \nu^2 \\
  b^{np,\chi}_{11} &= 9 \nu X_1 - \frac{15}{4} \nu^2 , \\
  b^{np,\chi}_{22} &= 9 \nu X_2 - \frac{15}{4} \nu^2 , \\
  b^{np,\chi}_{12} &= b^{np,\chi}_{21} = 3 \nu + \frac{9}{4} \nu^2 , \\
  b^{np,n\chi}_{11} &= 15 \nu X_1 - \frac{15}{4} \nu^2 , \\
  b^{np,n\chi}_{22} &= 15 \nu X_2 - \frac{15}{4} \nu^2 , \\
  b^{np,n\chi}_{12} &= b^{np,n\chi}_{21} = \frac{45}{8} \nu + \frac{15}{4} \nu^2 ,
\end{align}
\end{subequations}
modifying Eqs.~(2.52a)--(2.53c) and (2.62) in Ref.~\cite{Balmelli:2015zsa}.
The coefficients $a^\chi_{ij}$ from Eq.~(2.61) in Ref.~\cite{Balmelli:2015zsa} remain unchanged.
Also the coefficients $b^{np,\chi}_{ij}$ are unchanged, but their defining
Eq.~(2.62) in Ref.~\cite{Balmelli:2015zsa} is no longer valid, so we listed them here explicitly for clarity.
This solution has the three additional terms with coefficients $b^{np,n\chi}_{11}$, $b^{np,n\chi}_{22}$, and $b^{np,n\chi}_{12}$, which vanish in Ref.~\cite{Balmelli:2015zsa}.
The six symmetries between the coefficients in Ref.~\cite{Balmelli:2015zsa} are also absent in our solution (assuming the gauge conditions $b^{p,\chi}_{ij} = 0$).

\section{Hamiltonian coefficients after matching to PN results}
\label{app:coeffs}
In this Appendix, we present the results of matching the SEOB Hamiltonians using the procedure described in Sec.~\ref{sec:matching}.
Here, we express the multipole constants as
\begin{equation}
\tilde{C}_{\left.i(\text{ES}^2\right)} \equiv C_{\left.i(\text{ES}^2\right)} - 1, 
\quad
\tilde{C}_{\left.i(\text{BS}^2\right)} \equiv C_{\left.i(\text{BS}^2\right)} - 1, 
\quad
\text{etc.}
\end{equation}
such that the black hole results are easily obtained by setting $\tilde{C}_{\dots} = 0$.

The expressions for the Hamiltonians and the potentials given in this Appendix are provided as Supplemental Material \cite{simpleSEOBanc} in the form of \texttt{Mathematica} files.

\begin{widetext}

\subsection{Coefficients of the {\BD} Hamiltonian}
\label{app:coeffsBD}
The spin-orbit and spin-cubed PN corrections in Eq.~\eqref{HSOBD} are given by
\begin{subequations}
\begin{align}
G_S &= 2 \bigg[
1 + \frac{1}{\hat{r}_c}\frac{5 \nu }{16} + \frac{27 \nu }{16}  \hat{p}_r^2 
+\frac{1}{\hat{r}_c^2} \left(\frac{41 \nu ^2}{256}+\frac{51 \nu }{8}\right)
+\frac{\hat{p}_r^2}{\hat{r}_c} \left(\frac{21 \nu }{4}-\frac{49 \nu ^2}{128}\right) 
+\left(\frac{169 \nu ^2}{256}-\frac{5 \nu }{16}\right) \hat{p}_r^4
\bigg]^{-1}, \\
G_{S^*} &= \frac{3}{2} \bigg[
1 
+ \frac{1}{\hat{r}_c} \left(\frac{\nu }{2}+\frac{3}{4}\right)
+ \left(\frac{3 \nu }{2}+\frac{5}{4}\right) \hat{p}_r^2
+ \frac{\hat{p}_r^2}{\hat{r}_c} \left(-\frac{7 \nu ^2}{8}+5 \nu -1\right)
+\frac{1}{\hat{r}_c^2} \left(\frac{3 \nu ^2}{8}+\frac{29 \nu }{4}+\frac{27}{16}\right) \nonumber\\
&\qquad\qquad
+\left(\frac{3 \nu ^2}{8}+\frac{25 \nu }{12}+\frac{5}{48}\right) \hat{p}_r^4
\bigg]^{-1}, \\
%%%%%%%%%%%%%%%%%%
G_{S^3} &= 
\frac{1}{\hat{r}_c} \bigg\lbrace
\left(\bm{n}\cdot \bm{\chi}_1\right)^2 \left[\left(5 \nu +(5 \nu -5) X_1\right) \tilde{C}_{\left.1(\text{BS}^3\right)}+\left(\frac{9 \nu  X_1}{4}-\frac{9 \nu ^2}{4}\right) \tilde{C}_{\left.1(\text{ES}^2\right)}-\frac{5 \nu ^2}{2}+2 \nu +\left(\frac{9 \nu }{2}-2\right) X_1\right] \nonumber\\
&\quad\qquad
+\bm{\chi}_1^2 \left[\left((1-\nu ) X_1-\nu \right) \tilde{C}_{\left.1(\text{BS}^3\right)}+\left(\frac{3 \nu ^2}{4}-\frac{3 \nu  X_1}{4}\right) \tilde{C}_{\left.1(\text{ES}^2\right)}-\frac{\nu ^2}{4}+\frac{\nu  X_1}{4}\right] \nonumber\\
&\quad\qquad
+\bm{n}\cdot \bm{\chi}_1 \, \bm{n}\cdot \bm{\chi}_2 \left[\left(-6 \nu ^2-\frac{15 \nu  X_1}{2}\right) \tilde{C}_{\left.1(\text{ES}^2\right)}-\frac{5 \nu ^2}{3}-\frac{2 \nu  X_1}{3}\right] \nonumber\\
&\quad\qquad
+\bm{\chi}_1\cdot \bm{\chi}_2 \left[\left(\frac{3 \nu ^2}{2}+\frac{3 \nu  X_1}{2}\right) \tilde{C}_{\left.1(\text{ES}^2\right)}+\frac{4 \nu ^2}{3}-\frac{5 \nu  X_1}{12}\right] 
+\bm{\chi}_2^2 \left[\left(\frac{3 \nu  X_2}{4}-\frac{3 \nu ^2}{4}\right) \tilde{C}_{\left.2(\text{ES}^2\right)}-\frac{13 \nu ^2}{12}+\frac{2 \nu  X_2}{3}\right] \nonumber\\
&\quad\qquad
+\left(\bm{n}\cdot \bm{\chi}_2\right)^2 \left[\left(\frac{3 \nu ^2}{4}-\frac{15 \nu  X_2}{4}\right) \tilde{C}_{\left.2(\text{ES}^2\right)}+\frac{25 \nu ^2}{6}-\frac{17 \nu  X_2}{6}\right] 
\bigg\rbrace \nonumber\\
&\quad
+\frac{\hat{L}^2}{\hat{r}^2}
\left[
\left(\bm{n}\cdot \bm{\chi}_1\right)^2 \left(\frac{\nu ^2}{2}-\frac{\nu  X_1}{2}\right) + \bm{n}\cdot \bm{\chi}_1 \, \bm{n}\cdot \bm{\chi}_2 \left(\nu ^2-\nu  X_1\right) +\left(\bm{n}\cdot \bm{\chi}_2\right)^2\left(\frac{\nu  X_2}{2}-\frac{3 \nu ^2}{2}\right) 
\right] \nonumber\\
&\quad
+\hat{p}_r^2
\left[
\left(\bm{n}\cdot \bm{\chi}_1\right)^2 \left(\frac{5 \nu  X_1}{2}-\frac{5 \nu ^2}{2}\right) +\bm{n}\cdot \bm{\chi}_1 \, \bm{n}\cdot \bm{\chi}_2 \left(3 \nu  X_1-\nu ^2\right) +\left(\bm{n}\cdot \bm{\chi}_2\right)^2 \left(\frac{7 \nu ^2}{2}-\frac{\nu  X_2}{2}\right) 
\right],\\
%%%%%%%%%%%%%%%%%%
\tilde{G}_{S^3} &= G_{S^3} \text{ with }  1 \leftrightarrow 2.
\end{align}
\end{subequations}

The spin-squared corrections in Eq.~\eqref{S2ansatz} are given by
\begin{subequations}
\begin{align}
A^{SS} &= 
\frac{1}{\hat{r}_c^3} 
\bm{\chi}_1^2 \left(\nu -X_1\right) \tilde{C}_{\left.1(\text{ES}^2\right)} \nonumber \\
&\quad 
+\frac{1}{\hat{r}_c^4} 
\bigg\lbrace\bm{\chi}_1^2 \left[\left(6 \nu +(2 \nu -6) X_1\right) \tilde{C}_{\left.1(\text{ES}^2\right)}-\frac{\nu ^2}{2}+3 \nu  X_1\right]
+\frac{1}{2}\bm{\chi}_1\cdot\bm{\chi}_2 \left(2 \nu -\nu ^2\right)
\bigg\rbrace \nonumber\\
&\quad 
+ \frac{1}{\hat{r}_c^5} 
\bigg\lbrace
\bm{\chi}_1^2 \left[\left(-\frac{207 \nu ^2}{28}+\frac{275 \nu }{14}+\left(\frac{533 \nu }{28}-\frac{275}{14}\right) X_1\right) \tilde{C}_{\left.1(\text{ES}^2\right)}+\frac{3 \nu ^3}{8}-\frac{157 \nu ^2}{8}+\left(\frac{123 \nu }{4}-\frac{45 \nu ^2}{8}\right) X_1\right] \nonumber\\
&\qquad\qquad 
+\frac{1}{2}\bm{\chi}_1\cdot\bm{\chi}_2 \left(\frac{3 \nu ^3}{4}+\frac{145 \nu ^2}{8}+\frac{25 \nu }{2}\right)
\bigg\rbrace + 1 \leftrightarrow 2 ,\\
%%%%%%%%%%%%%%%%%%%%%%%
A^{nS} &= 
\frac{1}{\hat{r}_c^3} 
\left(\bm{n}\cdot\bm{\chi}_1\right)^2 \left(3 X_1-3 \nu \right) \tilde{C}_{\left.1(\text{ES}^2\right)}  \nonumber\\
&\quad 
+\frac{1}{\hat{r}_c^4} \bigg\lbrace
\left(\bm{n}\cdot\bm{\chi}_1\right)^2 \left[\left(-3 \nu ^2-9 \nu +(9-3 \nu ) X_1\right) \tilde{C}_{\left.1(\text{ES}^2\right)}-\frac{5 \nu ^2}{4}-7 \nu  X_1\right] 
+\frac{1}{2}(\bm{n}\cdot\bm{\chi}_2) (\bm{n}\cdot\bm{\chi}_1)\left(\frac{9 \nu ^2}{2}-\frac{27 \nu }{4}\right)
\bigg\rbrace \nonumber\\
&\quad 
+\frac{1}{\hat{r}_c^5} \bigg\lbrace
\left(\bm{n}\cdot\bm{\chi}_1\right)^2 \bigg[\left(-\frac{7 \nu ^3}{8}-\frac{641 \nu ^2}{56}-\frac{150 \nu }{7}+\left(-\frac{47 \nu ^2}{8}+\frac{22 \nu }{7}+\frac{150}{7}\right) X_1\right) \tilde{C}_{\left.1(\text{ES}^2\right)} +\frac{11 \nu ^3}{4}-\frac{71 \nu ^2}{12} \nonumber\\
&\qquad\qquad\quad
+\left(-\frac{63 \nu ^2}{4}-\frac{79 \nu }{3}\right) X_1\bigg] 
+\frac{1}{2}(\bm{n}\cdot\bm{\chi}_2) (\bm{n}\cdot\bm{\chi}_1) \left(\frac{3 \nu ^3}{2}-\frac{265 \nu ^2}{6}-\frac{387 \nu }{16}\right) 
\bigg\rbrace 
+ 1 \leftrightarrow 2, \\
%%%%%%%%%%%%%%%%%%%%
B_p^{nS} &= 
\frac{1}{\hat{r}_c^3} \bigg\lbrace
\left(\bm{n}\cdot \bm{\chi}_1\right)^2 \left[\left(-3 \nu ^2+3 \nu +(3 \nu -3) X_1\right) \tilde{C}_{\left.1(\text{ES}^2\right)}-\frac{5 \nu ^2}{2}+4 \nu  X_1\right] 
+ \frac{1}{2} \bm{n}\cdot \bm{\chi}_1 \, \bm{n}\cdot \bm{\chi}_2 \left(2 \nu ^2+\frac{9 \nu }{4}\right)
\bigg\rbrace \nonumber\\
&\quad
+\frac{1}{\hat{r}_c^4} \bigg\lbrace
\left(\bm{n}\cdot \bm{\chi}_1\right)^2 \bigg[\left(-\frac{7 \nu ^3}{8}-\frac{221 \nu ^2}{8}+\frac{15 \nu }{2}+\left(-\frac{47 \nu ^2}{8}+\frac{169 \nu }{4}-\frac{15}{2}\right) X_1\right) \tilde{C}_{\left.1(\text{ES}^2\right)} -\frac{889 \nu ^2}{24} +\frac{27 \nu ^3}{8} \nonumber\\
&\qquad\qquad\quad
+\left(\frac{323 \nu }{12}-\frac{217 \nu ^2}{8}\right) X_1\bigg] 
+\frac{1}{2}\bm{n}\cdot \bm{\chi}_1 \, \bm{n}\cdot \bm{\chi}_2 \left(\frac{11 \nu ^3}{4}-\frac{427 \nu ^2}{24}+\frac{57 \nu }{16}\right) 
\bigg\rbrace
+ 1 \leftrightarrow 2,
\\
%%%%%%%%%%%%%%%%%%%%
B_{np}^{nS} &=
\frac{1}{\hat{r}_c^3} 
\bigg\lbrace
\left(\bm{n}\cdot \bm{\chi}_1\right)^2 \left(\frac{15 \nu ^2}{4}-15 \nu  X_1\right)
+ \frac{1}{2}\bm{n}\cdot \bm{\chi}_1 \, \bm{n}\cdot \bm{\chi}_2 \left(-\frac{15 \nu ^2}{2}-\frac{45 \nu }{4}\right)
\bigg\rbrace
\nonumber\\
&\quad 
+\frac{1}{\hat{r}_c^4}
\bigg\lbrace
\left(\bm{n}\cdot \bm{\chi}_1\right)^2 \bigg[\left(-\frac{7 \nu ^3}{2}-\frac{\nu ^2}{4}+\frac{9 \nu }{2}+\left(\frac{25 \nu ^2}{2}+\frac{121 \nu }{4}-\frac{9}{2}\right) X_1\right) \tilde{C}_{\left.1(\text{ES}^2\right)} +\frac{17 \nu ^3}{8}+\frac{185 \nu ^2}{24}
\nonumber\\
&\qquad\qquad\quad 
+\left(-\frac{23 \nu ^2}{8}-\frac{619 \nu }{12}\right) X_1\bigg]
+\frac{1}{2}\bm{n}\cdot \bm{\chi}_1 \, \bm{n}\cdot \bm{\chi}_2 \left(-\frac{47 \nu ^3}{4}-\frac{4411 \nu ^2}{24}-\frac{39 \nu }{2}\right) 
\bigg\rbrace
+ 1 \leftrightarrow 2,
\\
%%%%%%%%%%%%%%%%%%%%
B_{np}^{SS} &=
\frac{1}{\hat{r}_c^3} \bigg\lbrace
\bm{\chi}_1^2 \left[\left(-3 \nu ^2+3 \nu +(3 \nu -3) X_1\right) \tilde{C}_{\left.1(\text{ES}^2\right)}-\frac{15 \nu ^2}{4}+9 \nu  X_1\right] 
+\frac{1}{2}\left(\frac{9 \nu ^2}{2}+6 \nu \right) \bm{\chi}_1\cdot \bm{\chi}_2
\bigg\rbrace \nonumber\\
&\quad
+\frac{1}{\hat{r}_c^4} \bigg\lbrace
\bm{\chi}_1^2 \left[\left(-\frac{159 \nu ^2}{4}+\frac{23 \nu }{2}+\left(-12 \nu ^2+\frac{197 \nu }{4}-\frac{23}{2}\right) X_1\right) \tilde{C}_{\left.1(\text{ES}^2\right)}+5 \nu ^3-61 \nu ^2+\left(\frac{275 \nu }{4}-37 \nu ^2\right) X_1\right] \nonumber\\
&\qquad\qquad
+\frac{1}{2}\left(10 \nu ^3+38 \nu ^2+20 \nu \right) \bm{\chi}_1\cdot \bm{\chi}_2
\bigg\rbrace
+ 1 \leftrightarrow 2,
\\
%%%%%%%%%%%%%%%%%%%%
Q^{S^2} &= 
\frac{\hat{p}_r^3}{\hat{r}_c^3}
\bigg\lbrace
\bm{n}\cdot \bm{\chi}_1 \, \hat{\bm{p}}\cdot \bm{\chi}_1 \left[\left(20 \nu ^3-35 \nu ^2+\left(35 \nu -20 \nu ^2\right) X_1\right) \tilde{C}_{\left.1(\text{ES}^2\right)}+\frac{199 \nu ^3}{8}-\frac{1085 \nu ^2}{24}+\left(\frac{130 \nu }{3}-\frac{517 \nu ^2}{8}\right) X_1\right] \nonumber\\
&\qquad\quad
+\bm{n}\cdot \bm{\chi}_1 \, \hat{\bm{p}}\cdot \bm{\chi}_2 \left(-\frac{79 \nu ^3}{8}+\frac{79 \nu ^2}{12}+\frac{45 \nu }{16}\right)
\bigg\rbrace \nonumber\\
&\quad 
+\frac{\hat{p}_r^4}{\hat{r}_c^3}
\bigg\lbrace
\bm{\chi}_1^2 \left[\left(5 \nu ^3-\frac{35 \nu ^2}{4}+\left(\frac{35 \nu }{4}-5 \nu ^2\right) X_1\right) \tilde{C}_{\left.1(\text{ES}^2\right)}+\frac{55 \nu ^3}{8}-\frac{105 \nu ^2}{8}+\left(\frac{55 \nu }{4}-\frac{145 \nu ^2}{8}\right) X_1\right]
 \nonumber\\
&\qquad\quad
+\left(\bm{n}\cdot \bm{\chi}_1\right)^2 \left[\left(\frac{245 \nu ^2}{4}-35 \nu ^3+\left(35 \nu ^2-\frac{245 \nu }{4}\right) X_1\right) \tilde{C}_{\left.1(\text{ES}^2\right)}-\frac{91 \nu ^3}{2}+\frac{1015 \nu ^2}{12}+\left(119 \nu ^2-\frac{1015 \nu }{12}\right) X_1\right]
 \nonumber\\
&\qquad\quad
+ \frac{1}{2}\bm{\chi}_1\cdot \bm{\chi}_2 \left(-\frac{25 \nu ^3}{4}+\frac{45 \nu ^2}{8}+\frac{5 \nu }{2}\right)
+ \frac{1}{2} \bm{n}\cdot \bm{\chi}_1 \, \bm{n}\cdot \bm{\chi}_2\left(\frac{77 \nu ^3}{2}-\frac{721 \nu ^2}{24}-\frac{105 \nu }{8}\right)
\bigg\rbrace
+ 1 \leftrightarrow 2.
\end{align}
\end{subequations}

The spin-quartic corrections in Eq.~\eqref{S4ansatz} are given by
\begin{align}
A^{S^4} &= \frac{1}{\hat{r}_c^5} \bigg\lbrace
\left(\bm{n}\cdot \bm{\chi}_1\right)^4
\left[\left(\frac{21 \nu ^2}{2}-\frac{21 \nu }{2}+\left(\frac{21}{2}-21 \nu \right) X_1\right) \tilde{C}_{\left.1(\text{ES}^2\right)}+\left(-\frac{35 \nu ^2}{4}+\frac{35 \nu }{4}+\left(\frac{35 \nu }{2}-\frac{35}{4}\right) X_1\right) \tilde{C}_{\left.1(\text{ES}^4\right)}\right]  \nonumber\\
&\qquad
+\bm{n}\cdot \bm{\chi}_2 \left(\bm{n}\cdot \bm{\chi}_1\right){}^3 \left[\left(35 \nu ^2-35 \nu  X_1\right) \tilde{C}_{\left.1(\text{BS}^3\right)}+\left(21 \nu  X_1-21 \nu ^2\right) \tilde{C}_{\left.1(\text{ES}^2\right)}\right] \nonumber\\
&\qquad
+\bm{\chi}_1\cdot \bm{\chi}_2 \left(\bm{n}\cdot \bm{\chi}_1\right)^2 \left[\left(15 \nu  X_1-15 \nu ^2\right) \tilde{C}_{\left.1(\text{BS}^3\right)}+\left(3 \nu ^2-3 \nu  X_1\right) \tilde{C}_{\left.1(\text{ES}^2\right)}\right] \nonumber\\
&\qquad
+\bm{\chi}_1^2 \left(\bm{n}\cdot \bm{\chi}_1\right)^2 \left[\left(-\frac{9 \nu ^2}{2}+\frac{9 \nu }{2}+\left(9 \nu -\frac{9}{2}\right) X_1\right) \tilde{C}_{\left.1(\text{ES}^2\right)}+\left(\frac{15 \nu ^2}{2}-\frac{15 \nu }{2}+\left(\frac{15}{2}-15 \nu \right) X_1\right) \tilde{C}_{\left.1(\text{ES}^4\right)}\right] \nonumber\\
&\qquad
+\bm{\chi}_2^2 \left(\bm{n}\cdot \bm{\chi}_1\right)^2 \left[6 \tilde{C}_{\left.2(\text{ES}^2\right)} \nu ^2+\tilde{C}_{\left.1(\text{ES}^2\right)} \left(\frac{15}{2} \tilde{C}_{\left.2(\text{ES}^2\right)} \nu ^2+\frac{15 \nu ^2}{2}\right)\right] \nonumber\\
&\qquad
+\frac{1}{2}\left(\bm{n}\cdot \bm{\chi}_2\right)^2 \left(\bm{n}\cdot \bm{\chi}_1\right)^2 \left[\tilde{C}_{\left.1(\text{ES}^2\right)} \left(-\frac{1}{2} 105 \tilde{C}_{\left.2(\text{ES}^2\right)} \nu ^2-42 \nu ^2\right)-42 \nu ^2 \tilde{C}_{\left.2(\text{ES}^2\right)}\right] \nonumber\\
&\qquad
+\bm{\chi}_1^2 \, \bm{n}\cdot \bm{\chi}_2 \, \bm{n}\cdot \bm{\chi}_1 \left[\left(15 \nu  X_1-15 \nu ^2\right) \tilde{C}_{\left.1(\text{BS}^3\right)}+\left(6 \nu ^2-6 \nu  X_1\right) \tilde{C}_{\left.1(\text{ES}^2\right)}\right] \nonumber\\
&\qquad
+\frac{1}{2}\bm{n}\cdot \bm{\chi}_2 \, \bm{\chi}_1\cdot \bm{\chi}_2 \, \bm{n}\cdot \bm{\chi}_1 \left[27 \tilde{C}_{\left.2(\text{ES}^2\right)} \nu ^2+\tilde{C}_{\left.1(\text{ES}^2\right)} \left(30 \tilde{C}_{\left.2(\text{ES}^2\right)} \nu ^2+27 \nu ^2\right)\right] \nonumber\\
&\qquad
+\bm{\chi}_1\cdot \bm{\chi}_2 \, \bm{\chi}_1^2 \left(3 \nu ^2-3 \nu  X_1\right)  \tilde{C}_{\left.1(\text{BS}^3\right)}
+\bm{\chi}_1^4 \left[-\frac{3 \nu ^2}{4}+\frac{3 \nu }{4}+\left(\frac{3 \nu }{2}-\frac{3}{4}\right) X_1\right]  \tilde{C}_{\left.1(\text{ES}^4\right)} \nonumber\\
&\qquad
+\frac{1}{2}\bm{\chi}_1^2 \bm{\chi}_2^2 \left[\tilde{C}_{\left.1(\text{ES}^2\right)} \left(-\frac{1}{2} 3 \tilde{C}_{\left.2(\text{ES}^2\right)} \nu ^2-\frac{3 \nu ^2}{2}\right)-\frac{3}{2} \nu ^2 \tilde{C}_{\left.2(\text{ES}^2\right)}\right]  \nonumber\\
&\qquad
+\frac{1}{2}\left(\bm{\chi}_1\cdot \bm{\chi}_2\right)^2 \left[\tilde{C}_{\left.1(\text{ES}^2\right)} \left(-3 \tilde{C}_{\left.2(\text{ES}^2\right)} \nu ^2-3 \nu ^2\right)-3 \nu ^2 \tilde{C}_{\left.2(\text{ES}^2\right)}\right]
\bigg\rbrace
+ 1 \leftrightarrow 2.
\end{align}

\subsection{Coefficients of the {\simp} Hamiltonian}
\label{app:coeffsSimple}

The spin-orbit and spin-cubed PN corrections in Eq.~\eqref{HSOsimp} are given by
\begin{subequations}
\begin{align}
G_S &= 2 \bigg[
1-\frac{27}{16} \nu  \hat{p}_r^2 -\frac{5 \nu }{16\hat{r}}
+\left(\frac{35 \nu ^2}{16}+\frac{5 \nu }{16}\right) \hat{p}_r^4
+\frac{\hat{p}_r^2}{\hat{r}}\left(\frac{23 \nu ^2}{16}-\frac{21 \nu }{4}\right)
+\frac{1}{\hat{r}^2}\left(-\frac{\nu ^2}{16}-\frac{51 \nu }{8}\right)
\bigg], \\
G_{S^*} &= \frac{3}{2} \bigg[
1 -\left(\frac{3 \nu }{2}+\frac{5}{4}\right) \hat{p}_r^2
-\frac{1}{\hat{r}}\left(\frac{3}{4}+\frac{\nu }{2}\right)
+\left(\frac{15 \nu ^2}{8}+\frac{5 \nu }{3}+\frac{35}{24}\right) \hat{p}_r^4
+\frac{\hat{p}_r^2}{\hat{r}} \left(\frac{19 \nu ^2}{8}-\frac{3 \nu }{2}+\frac{23}{8}\right) \nonumber\\
&\quad\qquad
+\frac{1}{\hat{r}^2} \left(-\frac{\nu ^2}{8}-\frac{13 \nu }{2}-\frac{9}{8}\right)
\bigg], \\
%%%%%%%%%%%%%%%%%%
G_{S^3} &= 
\frac{1}{\hat{r}} \bigg\lbrace
\left(\bm{n}\cdot \bm{\chi}_1\right)^2 \left[\left(5 \nu +(5 \nu -5) X_1\right) \tilde{C}_{\left.1(\text{BS}^3\right)}+\left(\frac{9 \nu  X_1}{4}-\frac{9 \nu ^2}{4}\right) \tilde{C}_{\left.1(\text{ES}^2\right)}-2 \nu ^2+2 \nu  X_1\right] \nonumber\\
&\quad\qquad
+\bm{\chi}_1^2 \left[\left((1-\nu ) X_1-\nu \right) \tilde{C}_{\left.1(\text{BS}^3\right)}+\left(\frac{3 \nu ^2}{4}-\frac{3 \nu  X_1}{4}\right) \tilde{C}_{\left.1(\text{ES}^2\right)}+\frac{\nu ^2}{2}-\frac{\nu  X_1}{2}\right] \nonumber\\
&\quad\qquad
+\bm{\chi}_1\cdot \bm{\chi}_2 \left[\left(\frac{3 \nu ^2}{2}+\frac{3 \nu  X_1}{2}\right) \tilde{C}_{\left.1(\text{ES}^2\right)}-\frac{\nu ^2}{6}-\frac{5 \nu  X_1}{12}\right]
+\bm{\chi}_2^2 \left[\left(\frac{3 \nu  X_2}{4}-\frac{3 \nu ^2}{4}\right) \tilde{C}_{\left.2(\text{ES}^2\right)}-\frac{\nu ^2}{3}-\frac{\nu  X_2}{12}\right] \nonumber\\
&\quad\qquad
+\bm{n}\cdot \bm{\chi}_1 \, \bm{n}\cdot \bm{\chi}_2 \left[\left(-6 \nu ^2-\frac{15 \nu  X_1}{2}\right) \tilde{C}_{\left.1(\text{ES}^2\right)}-\frac{8 \nu ^2}{3}+\frac{10 \nu  X_1}{3}\right] \nonumber\\
&\quad\qquad
+\left(\bm{n}\cdot \bm{\chi}_2\right)^2 \left[\left(\frac{3 \nu ^2}{4}-\frac{15 \nu  X_2}{4}\right) \tilde{C}_{\left.2(\text{ES}^2\right)}+\frac{14 \nu ^2}{3}-\frac{4 \nu  X_2}{3}\right] 
\bigg\rbrace \nonumber\\
&\quad
+\frac{\hat{L}^2}{r^2}
\left[
\left(\bm{n}\cdot \bm{\chi}_1\right)^2 \left(\frac{\nu ^2}{2}-\frac{\nu  X_1}{2}\right) + \bm{n}\cdot \bm{\chi}_1 \, \bm{n}\cdot \bm{\chi}_2\left(\nu ^2-\nu  X_1\right) + \left(\bm{n}\cdot \bm{\chi}_2\right)^2\left(\frac{\nu  X_2}{2}-\frac{3 \nu ^2}{2}\right) 
\right] \nonumber\\
&\quad
+\hat{p}_r^2
\left[
\left(\bm{n}\cdot \bm{\chi}_1\right)^2 \left(\frac{5 \nu  X_1}{2}-\frac{5 \nu ^2}{2}\right) +\bm{n}\cdot \bm{\chi}_1 \, \bm{n}\cdot \bm{\chi}_2 \left(3 \nu  X_1-\nu ^2\right) +\left(\bm{n}\cdot \bm{\chi}_2\right)^2\left(\frac{7 \nu ^2}{2}-\frac{\nu  X_2}{2}\right) 
\right],\\
%%%%%%%%%%%%%%%%%%
\tilde{G}_{S^3} &= G_{S^3} \text{ with }  1 \leftrightarrow 2.
\end{align}
\end{subequations}

The spin-squared and spin-quartic corrections in Eq.~\eqref{S2S4simp} read 
\begin{subequations}
\begin{align}
A^{SS} &= 
\frac{1}{\hat{r}^3} 
\bm{\chi}_1^2 \left(\nu -X_1\right) \tilde{C}_{\left.1(\text{ES}^2\right)} \nonumber \\
&\quad 
+\frac{1}{\hat{r}^4} 
\bigg\lbrace
\bm{\chi}_1^2 \left[\left(4 \nu +(2 \nu -4) X_1\right) \tilde{C}_{\left.1(\text{ES}^2\right)}-\frac{\nu ^2}{2}+3 \nu  X_1\right]
+\frac{1}{2}\bm{\chi}_1\cdot \bm{\chi}_2\left(2 \nu -\nu ^2\right)
\bigg\rbrace \nonumber\\
&\quad 
+ \frac{1}{\hat{r}^5} 
\bigg\lbrace
\bm{\chi}_1^2 \left[\left(-\frac{207 \nu ^2}{28}+\frac{163 \nu }{14}+\left(\frac{421 \nu }{28}-\frac{163}{14}\right) X_1\right) \tilde{C}_{\left.1(\text{ES}^2\right)}+\frac{3 \nu ^3}{8}-\frac{125 \nu ^2}{8}+\left(\frac{87 \nu }{4}-\frac{45 \nu ^2}{8}\right) X_1\right] \nonumber\\
&\qquad\qquad
+\frac{1}{2}\bm{\chi}_1\cdot \bm{\chi}_2\left(\frac{3 \nu ^3}{4}+\frac{113 \nu ^2}{8}+\frac{17 \nu }{2}\right)
\bigg\rbrace + 1 \leftrightarrow 2,\\
%%%%%%%%%%%%%%%%%%%%%%%
A^{nS} &= 
\frac{1}{\hat{r}^3} 
\left(\bm{n}\cdot \bm{\chi}_1\right)^2\left(3 X_1-3 \nu \right) \tilde{C}_{\left.1(\text{ES}^2\right)} 
 \nonumber\\
&\quad 
+\frac{1}{\hat{r}^4} \bigg\lbrace
\left(\bm{n}\cdot \bm{\chi}_1\right)^2 \left[\left(-3 \nu ^2-9 \nu +(9-3 \nu ) X_1\right) \tilde{C}_{\left.1(\text{ES}^2\right)}-\frac{5 \nu ^2}{4}-7 \nu  X_1\right] 
+\frac{1}{2}\bm{n}\cdot \bm{\chi}_1 \, \bm{n}\cdot \bm{\chi}_2 \left(\frac{9 \nu ^2}{2}-\frac{27 \nu }{4}\right) 
\bigg\rbrace \nonumber\\
&\quad 
+\frac{1}{\hat{r}^5} \bigg\lbrace
\left(\bm{n}\cdot \bm{\chi}_1\right)^2 \bigg[\left(-\frac{7 \nu ^3}{8}-\frac{641 \nu ^2}{56}-\frac{150 \nu }{7}+\left(-\frac{47 \nu ^2}{8}+\frac{22 \nu }{7}+\frac{150}{7}\right) X_1\right) \tilde{C}_{\left.1(\text{ES}^2\right)}+\frac{11 \nu ^3}{4}-\frac{71 \nu ^2}{12} \nonumber\\
&\qquad\qquad\quad
+\left(-\frac{63 \nu ^2}{4}-\frac{79 \nu }{3}\right) X_1\bigg] 
+\frac{1}{2}\bm{n}\cdot \bm{\chi}_1 \, \bm{n}\cdot \bm{\chi}_2\left(\frac{3 \nu ^3}{2}-\frac{265 \nu ^2}{6}-\frac{387 \nu }{16}\right) 
\bigg\rbrace + 1 \leftrightarrow 2, \\
%%%%%%%%%%%%%%%%%%%%
B_p^{nS} &= 
\frac{1}{\hat{r}^3} \bigg\lbrace
\left(\bm{n}\cdot \bm{\chi}_1\right)^2 \left[\left(3 \nu ^2-3 \nu +(3-3 \nu ) X_1\right) \tilde{C}_{\left.1(\text{ES}^2\right)}+\frac{5 \nu ^2}{2}-4 \nu  X_1\right] 
+\frac{1}{2}\bm{n}\cdot \bm{\chi}_1 \, \bm{n}\cdot \bm{\chi}_2\left(-2 \nu ^2-\frac{9 \nu }{4}\right) 
\bigg\rbrace \nonumber\\
&\quad
+\frac{1}{\hat{r}^4} \bigg\lbrace
\left(\bm{n}\cdot \bm{\chi}_1\right)^2 \bigg[\left(\frac{7 \nu ^3}{8}+\frac{221 \nu ^2}{8}-\frac{15 \nu }{2}+\left(\frac{47 \nu ^2}{8}-\frac{169 \nu }{4}+\frac{15}{2}\right) X_1\right) \tilde{C}_{\left.1(\text{ES}^2\right)} -\frac{27 \nu ^3}{8}+\frac{889 \nu ^2}{24} \nonumber\\
&\qquad\qquad\quad
+\left(\frac{217 \nu ^2}{8}-\frac{323 \nu }{12}\right) X_1\bigg] 
+\frac{1}{2}\bm{n}\cdot \bm{\chi}_1 \, \bm{n}\cdot \bm{\chi}_2\left(-\frac{11 \nu ^3}{4}+\frac{427 \nu ^2}{24}-\frac{57 \nu }{16}\right) 
\bigg\rbrace + 1 \leftrightarrow 2,
\\
%%%%%%%%%%%%%%%%%%%%
B_{np}^{nS} &=
\frac{1}{\hat{r}^3} 
\bigg\lbrace
\left(\bm{n}\cdot \bm{\chi}_1\right)^2\left(\frac{15 \nu ^2}{4}-15 \nu  X_1\right) 
+\frac{1}{2}\bm{n}\cdot \bm{\chi}_1 \, \bm{n}\cdot \bm{\chi}_2 \left(-\frac{15 \nu ^2}{2}-\frac{45 \nu }{4}\right) 
\bigg\rbrace
\nonumber\\
&\quad 
+\frac{1}{\hat{r}^4}
\bigg\lbrace
\left(\bm{n}\cdot \bm{\chi}_1\right)^2 \bigg[\left(-\frac{7 \nu ^3}{2}-\frac{\nu ^2}{4}+\frac{9 \nu }{2}+\left(\frac{25 \nu ^2}{2}+\frac{121 \nu }{4}-\frac{9}{2}\right) X_1\right) \tilde{C}_{\left.1(\text{ES}^2\right)} +\frac{17 \nu ^3}{8}+\frac{365 \nu ^2}{24}\nonumber\\
&\qquad\qquad\quad
+\left(-\frac{23 \nu ^2}{8}-\frac{979 \nu }{12}\right) X_1\bigg]
+\frac{1}{2}\left(-\frac{47 \nu ^3}{4}-\frac{4771 \nu ^2}{24}-42 \nu \right) \bm{n}\cdot \bm{\chi}_1 \, \bm{n}\cdot \bm{\chi}_2
\bigg\rbrace + 1 \leftrightarrow 2,
\\
%%%%%%%%%%%%%%%%%%%%
B_{np}^{SS} &=
\frac{1}{\hat{r}^3} \bigg\lbrace
\bm{\chi}_1^2 \left[\left(-3 \nu ^2+3 \nu +(3 \nu -3) X_1\right) \tilde{C}_{\left.1(\text{ES}^2\right)}-\frac{15 \nu ^2}{4}+9 \nu  X_1\right] 
+\frac{1}{2}\bm{\chi}_1\cdot \bm{\chi}_2\left(\frac{9 \nu ^2}{2}+6 \nu \right) 
\bigg\rbrace \nonumber\\
&\quad
+\frac{1}{\hat{r}^4} \bigg\lbrace
\bm{\chi}_1^2 \left[\left(-\frac{159 \nu ^2}{4}+\frac{23 \nu }{2}+\left(-12 \nu ^2+\frac{197 \nu }{4}-\frac{23}{2}\right) X_1\right) \tilde{C}_{\left.1(\text{ES}^2\right)}+5 \nu ^3-55 \nu ^2+\left(\frac{251 \nu }{4}-37 \nu ^2\right) X_1\right] \nonumber\\
&\qquad\qquad
+\frac{1}{2}\bm{\chi}_1\cdot \bm{\chi}_2 \left(10 \nu ^3+26 \nu ^2+20 \nu \right) 
\bigg\rbrace + 1 \leftrightarrow 2,
\\
%%%%%%%%%%%%%%%%%%%%
Q^{S^2} &= 
\frac{\hat{p}_r^3}{\hat{r}^3}
\bigg\lbrace
\bm{n}\cdot \bm{\chi}_1 \, \hat{\bm{p}}\cdot \bm{\chi}_1 \left[\left(20 \nu ^3-35 \nu ^2+\left(35 \nu -20 \nu ^2\right) X_1\right) \tilde{C}_{\left.1(\text{ES}^2\right)}+\frac{199 \nu ^3}{8}-\frac{1085 \nu ^2}{24}+\left(\frac{130 \nu }{3}-\frac{517 \nu ^2}{8}\right) X_1\right] \nonumber\\
&\qquad
+ \bm{n}\cdot \bm{\chi}_1 \, \hat{\bm{p}}\cdot \bm{\chi}_2 \left(-\frac{79 \nu ^3}{8}+\frac{79 \nu ^2}{12}+\frac{45 \nu }{16}\right) 
\bigg\rbrace \nonumber\\
&\quad 
+\frac{\hat{p}_r^4}{\hat{r}^3}
\bigg\lbrace
\bm{\chi}_1^2 \left[\left(5 \nu ^3-\frac{35 \nu ^2}{4}+\left(\frac{35 \nu }{4}-5 \nu ^2\right) X_1\right) \tilde{C}_{\left.1(\text{ES}^2\right)}+\frac{55 \nu ^3}{8}-\frac{105 \nu ^2}{8}+\left(\frac{55 \nu }{4}-\frac{145 \nu ^2}{8}\right) X_1\right] \nonumber\\
&\qquad
+\left(\bm{n}\cdot \bm{\chi}_1\right)^2 \left[\left(\frac{245 \nu ^2}{4}-35 \nu ^3+\left(35 \nu ^2-\frac{245 \nu }{4}\right) X_1\right) \tilde{C}_{\left.1(\text{ES}^2\right)}-\frac{91 \nu ^3}{2}+\frac{1015 \nu ^2}{12}+\left(119 \nu ^2-\frac{1015 \nu }{12}\right) X_1\right]
 \nonumber\\
&\qquad
+\frac{1}{2}\bm{\chi}_1\cdot \bm{\chi}_2\left(-\frac{25 \nu ^3}{4}+\frac{45 \nu ^2}{8}+\frac{5 \nu }{2}\right) 
+\frac{1}{2}\bm{n}\cdot \bm{\chi}_1 \, \bm{n}\cdot \bm{\chi}_2\left(\frac{77 \nu ^3}{2}-\frac{721 \nu ^2}{24}-\frac{105 \nu }{8}\right) 
\bigg\rbrace + 1 \leftrightarrow 2, \\
%%%%%%%%%%%%%%%%%%%%%%%%%%%%
A^{S^4} &= \frac{1}{\hat{r}^5} \bigg\lbrace
\left(\bm{n}\cdot \bm{\chi}_1\right)^4 \left[\left(\frac{15 \nu ^2}{2}-\frac{15 \nu }{2}+\left(\frac{15}{2}-15 \nu \right) X_1\right) \tilde{C}_{\left.1(\text{ES}^2\right)}+\left(\frac{35 \nu }{4}-\frac{35 \nu ^2}{4}+\left(\frac{35 \nu }{2}-\frac{35}{4}\right) X_1\right) \tilde{C}_{\left.1(\text{ES}^4\right)}\right] \nonumber\\
&\qquad
+\bm{n}\cdot \bm{\chi}_2 \left(\bm{n}\cdot \bm{\chi}_1\right){}^3 \left[\left(35 \nu ^2-35 \nu  X_1\right) \tilde{C}_{\left.1(\text{BS}^3\right)}+\left(15 \nu  X_1-15 \nu ^2\right) \tilde{C}_{\left.1(\text{ES}^2\right)}\right] \nonumber\\
&\qquad
+\bm{\chi}_1\cdot \bm{\chi}_2 \left(\bm{n}\cdot \bm{\chi}_1\right)^2
\left[\left(15 \nu  X_1-15 \nu ^2\right) \tilde{C}_{\left.1(\text{BS}^3\right)}+\left(12 \nu ^2-12 \nu  X_1\right) \tilde{C}_{\left.1(\text{ES}^2\right)}\right] \nonumber\\
&\qquad
+\bm{\chi}_1^2 \left(\bm{n}\cdot \bm{\chi}_1\right)^2\left[\left(-9 \nu ^2+9 \nu +(18 \nu -9) X_1\right) \tilde{C}_{\left.1(\text{ES}^2\right)}+\left(\frac{15 \nu ^2}{2}-\frac{15 \nu }{2}+\left(\frac{15}{2}-15 \nu \right) X_1\right) \tilde{C}_{\left.1(\text{ES}^4\right)}\right] \nonumber\\
&\qquad
+\bm{\chi}_2^2 \left(\bm{n}\cdot \bm{\chi}_1\right)^2 \left[6 \tilde{C}_{\left.2(\text{ES}^2\right)} \nu ^2+\tilde{C}_{\left.1(\text{ES}^2\right)} \left(\frac{15}{2} \tilde{C}_{\left.2(\text{ES}^2\right)} \nu ^2+3 \nu ^2\right)\right] \nonumber\\
&\qquad
+\frac{1}{2}\left(\bm{n}\cdot \bm{\chi}_2\right)^2 \left(\bm{n}\cdot \bm{\chi}_1\right)^2 \left[\tilde{C}_{\left.1(\text{ES}^2\right)} \left(-\frac{1}{2} 105 \tilde{C}_{\left.2(\text{ES}^2\right)} \nu ^2-45 \nu ^2\right)-45 \nu ^2 \tilde{C}_{\left.2(\text{ES}^2\right)}\right] \nonumber\\
&\qquad
+\bm{n}\cdot \bm{\chi}_2 \, \bm{\chi}_1^2 \bm{n}\cdot \bm{\chi}_1 \left[\left(15 \nu  X_1-15 \nu ^2\right) \tilde{C}_{\left.1(\text{BS}^3\right)}+\left(6 \nu ^2-6 \nu  X_1\right) \tilde{C}_{\left.1(\text{ES}^2\right)}\right] \nonumber\\
&\qquad
+\frac{1}{2}\bm{n}\cdot \bm{\chi}_2 \, \bm{\chi}_1\cdot \bm{\chi}_2 \, \bm{n}\cdot \bm{\chi}_1 \left[27 \tilde{C}_{\left.2(\text{ES}^2\right)} \nu ^2+\tilde{C}_{\left.1(\text{ES}^2\right)} \left(30 \tilde{C}_{\left.2(\text{ES}^2\right)} \nu ^2+27 \nu ^2\right)\right] \nonumber\\
&\qquad
+\bm{\chi}_1\cdot \bm{\chi}_2 \, \bm{\chi}_1^2 \left[\left(3 \nu ^2-3 \nu  X_1\right) \tilde{C}_{\left.1(\text{BS}^3\right)}+\left(3 \nu  X_1-3 \nu ^2\right) \tilde{C}_{\left.1(\text{ES}^2\right)}\right] 
-\frac{3}{4} \nu ^2 \bm{\chi}_1^2 \bm{\chi}_2^2 \tilde{C}_{\left.1(\text{ES}^2\right)} \tilde{C}_{\left.2(\text{ES}^2\right)} \nonumber\\
&\qquad
+\bm{\chi}_1^4 \left[\left(\frac{3 \nu ^2}{2}-\frac{3 \nu }{2}+\left(\frac{3}{2}-3 \nu \right) X_1\right) \tilde{C}_{\left.1(\text{ES}^2\right)}+\left(-\frac{3 \nu ^2}{4}+\frac{3 \nu }{4}+\left(\frac{3 \nu }{2}-\frac{3}{4}\right) X_1\right) \tilde{C}_{\left.1(\text{ES}^4\right)}\right] \nonumber\\
&\qquad
+\frac{1}{2}\left(\bm{\chi}_1\cdot \bm{\chi}_2\right)^2 \left[\tilde{C}_{\left.1(\text{ES}^2\right)} \left(-3 \tilde{C}_{\left.2(\text{ES}^2\right)} \nu ^2-3 \nu ^2\right)-3 \nu ^2 \tilde{C}_{\left.2(\text{ES}^2\right)}\right]
\bigg\rbrace + 1 \leftrightarrow 2.
\end{align}
\end{subequations}

\subsection{Coefficients of the {\DN} Hamiltonian}
\label{app:coeffsDN}
The coefficients of the {\DN} Hamiltonian are given in Ref.~\cite{Nagar:2018plt}, but we rewrite them here for convenience in the notation used in the rest of the paper.

The spin-orbit and spin-cubic correction in Eq.~\eqref{SOS3DN} are given by
\begin{subequations}
\begin{align}\label{inversetaylorG}
G_S &= 2 \bigg[
1 +\frac{27}{16} \nu  \hat{p}_{r_\ast}^2 +\frac{5 \nu }{16 \hat{r}_c}
+\left(\frac{169 \nu ^2}{256}-\frac{5 \nu }{16}\right) \hat{p}_{r_\ast}^4 
+\frac{\hat{p}_{r_\ast}^2}{\hat{r}_c} \left(12 \nu -\frac{49 \nu ^2}{128}\right)
+\frac{1}{\hat{r}_c^2} \left(\frac{41 \nu ^2}{256}+\frac{51 \nu }{8}\right)
\bigg]^{-1}, \nonumber\\
G_{S^\ast} &= \frac{3}{2} \bigg[
1 +\frac{1}{\hat{r}_c} \left(\frac{\nu }{2}+\frac{3}{4}\right)
+\left(\frac{3 \nu }{2}+\frac{5}{4}\right) \hat{p}_{r_\ast}^2 
+\left(\frac{3 \nu ^2}{8}+\frac{25 \nu }{12}+\frac{5}{48}\right) \hat{p}_{r_\ast}^4
+ \frac{\hat{p}_{r_\ast}^2}{\hat{r}_c} \left(-\frac{7 \nu ^2}{8}+11 \nu +4\right) \nonumber\\
&\qquad
+\frac{1}{\hat{r}_c^2} \left(\frac{3 \nu ^2}{8}+\frac{29 \nu }{4}+\frac{27}{16}\right)
\bigg]^{-1}, \\
%%%%%%%%%%%%%%%%%%%%%%%%%%%
G_{S^3} &= \frac{1}{\hat{r}_c} \bigg\lbrace
\chi _1^2 \bigg[\left(1-2 \nu +(\nu -1) X_1\right) \tilde{C}_{\left.1(\text{BS}^3\right)}+\left(\frac{\nu ^2}{4}+\frac{5 \nu }{4}+\left(\frac{3}{4}-\frac{\nu }{2}\right) X_1-\frac{3}{4}\right) \tilde{C}_{\left.1(\text{ES}^2\right)}+\frac{\nu ^2}{4}-\frac{3 \nu }{4}+\frac{1}{4} \nonumber\\
&\qquad\qquad
+\left(\frac{\nu }{2}-\frac{1}{4}\right) X_1\bigg] 
+\chi_1\chi_2 \left[\left(\frac{\nu ^2}{4}+2 \nu -2 \nu  X_1\right) \tilde{C}_{\left.1(\text{ES}^2\right)}-\frac{\nu ^2}{4}+\frac{\nu }{2}-\frac{\nu  X_1}{2}\right]
\bigg\rbrace, \nonumber\\
%%%%%%%%%%%%%%%%%%%%%%%
\tilde{G}_{S^3} &= G_{S^3} \text{ with }  1 \leftrightarrow 2.
\end{align}
\end{subequations}

The spin-squared and spin-quartic corrections in Eq.~\eqref{S2S4DN} are given by
\begin{subequations}
\begin{align}
\delta a^2_\text{NLO} &= \chi_1^2 \left[\left((4-2 \nu ) X_1-4 \nu \right) \tilde{C}_{\left.1(\text{ES}^2\right)}+\frac{\nu ^2}{2}-3 \nu  X_1\right]  
+\frac{1}{2}\chi_1\chi_2\left(\nu ^2-2 \nu \right) + 1 \leftrightarrow 2,  \\
%%%%%%%%%%%%%%%%%%%%
\delta a^2_\text{NNLO} &= 
\chi_1^2 \left[\left(\frac{207 \nu ^2}{28}-\frac{275 \nu }{14}+\left(\frac{275}{14}-\frac{533 \nu }{28}\right) X_1\right) \tilde{C}_{\left.1(\text{ES}^2\right)}-\frac{3 \nu ^3}{8}+\frac{157 \nu ^2}{8}+\left(\frac{45 \nu ^2}{8}-\frac{123 \nu }{4}\right) X_1\right] \nonumber\\
&\quad 
+\frac{1}{2}\chi_1\chi_2\left(-\frac{3 \nu ^3}{4}-\frac{145 \nu ^2}{8}-\frac{25 \nu }{2}\right) + 1 \leftrightarrow 2, \\
%%%%%%%%%%%%%%%%%%%%
\delta a^4_\text{LO} &= 
\chi _1^3 \chi _2 \left[\left(3 \nu  X_1-3 \nu ^2\right) \tilde{C}_{\left.1(\text{BS}^3\right)}+\left(3 \nu ^2-3 \nu  X_1\right) \tilde{C}_{\left.1(\text{ES}^2\right)}\right]  \nonumber\\
&\quad 
+ \frac{1}{2}\chi_1^2\chi_2^2 \left[\tilde{C}_{\left.1(\text{ES}^2\right)} \left(3 \nu ^2 \tilde{C}_{\left.2(\text{ES}^2\right)}+3 \nu ^2\right)+3 \nu ^2 \tilde{C}_{\left.2(\text{ES}^2\right)}\right]\nonumber\\
&\quad 
+\chi_1^4 \bigg[\left(-\frac{3 \nu ^2}{4}+\frac{9 \nu }{4}+\left(\frac{3}{4}-\frac{3 \nu }{2}\right) X_2-\frac{3}{4}\right) \tilde{C}_{\left.1(\text{ES}^2\right)}^2+\left(-\frac{3 \nu ^2}{2}+\frac{9 \nu }{2}+\left(\frac{3}{2}-3 \nu \right) X_2-\frac{3}{2}\right) \tilde{C}_{\left.1(\text{ES}^2\right)} \nonumber\\
&\qquad\qquad
+\left(\frac{3 \nu ^2}{4}-\frac{9 \nu }{4}+\left(\frac{3 \nu }{2}-\frac{3}{4}\right) X_2+\frac{3}{4}\right) \tilde{C}_{\left.1(\text{ES}^4\right)}\bigg] 
 + 1 \leftrightarrow 2.
\end{align}
\end{subequations}

\subsection{Coefficients of the {\BB} Hamiltonian}
\label{app:coeffsBB}

The spin-squared corrections in Eq.~\eqref{HBBansatz} are given by
\begin{subequations}
\begin{align}
A^{SS} &= 
\frac{1}{\hat{r}^3} 
\bm{\chi}_1^2 \left(\nu -X_1\right) \tilde{C}_{\left.1(\text{ES}^2\right)}
\nonumber \\
&\quad 
+ \frac{1}{\hat{r}^4} 
\bigg\lbrace
\bm{\chi}_1^2 \left[\left(2 \nu +(2 \nu -2) X_1\right) \tilde{C}_{\left.1(\text{ES}^2\right)}+\frac{17 \nu ^2}{6}+\left(\frac{23 \nu ^2}{6}+\frac{\nu }{6}\right) X_1\right]  
+\frac{37}{12} \nu ^2 \bm{\chi}_1\cdot \bm{\chi}_2 
\bigg\rbrace \nonumber\\
&\quad 
+ \frac{1}{\hat{r}^5} 
\bigg\lbrace
\bm{\chi}_1^2 \bigg[\left(-\frac{207 \nu ^2}{28}+\frac{51 \nu }{14}+\left(\frac{309 \nu }{28}-\frac{51}{14}\right) X_1\right) \tilde{C}_{\left.1(\text{ES}^2\right)}
+\frac{529 \nu ^4}{144}-\frac{2353 \nu ^3}{144}+\frac{55 \nu ^2}{4} \nonumber\\
&\qquad\qquad\quad
+\left(\frac{143 \nu ^3}{72}+\frac{7015 \nu ^2}{144}-\frac{155 \nu }{8}\right) X_1\bigg]
+\frac{1}{2}\bm{\chi}_1\cdot \bm{\chi}_2\left(\frac{529 \nu ^4}{72}-\frac{112 \nu ^3}{3}+\frac{51 \nu ^2}{8}\right)  
\bigg\rbrace  
+ 1 \leftrightarrow 2,\\
%%%%%%%%%%%%%%%%%%%%%%%
A^{nS} &= 
\frac{1}{\hat{r}^3} 
\left(3 X_1-3 \nu \right) \tilde{C}_{\left.1(\text{ES}^2\right)} \left(\bm{n}\cdot \bm{\chi}_1\right)^2 \nonumber\\
&\quad 
+\frac{1}{\hat{r}^4} \bigg\lbrace
\left(\bm{n}\cdot \bm{\chi}_1\right)^2 \left[\left(-3 \nu ^2-3 \nu +(3-3 \nu ) X_1\right) \tilde{C}_{\left.1(\text{ES}^2\right)}-\frac{5 \nu ^3}{4}+\frac{3 \nu ^2}{2}+\left(-\frac{15 \nu ^2}{4}-6 \nu \right) X_1\right] \nonumber\\
&\qquad\qquad
+\frac{1}{2}\bm{n}\cdot \bm{\chi}_1 \, \bm{n}\cdot \bm{\chi}_2 \left(-\frac{5 \nu ^3}{2}-\frac{67 \nu ^2}{4}\right) 
\bigg\rbrace \nonumber\\
&\quad 
+\frac{1}{\hat{r}^5} \bigg\lbrace
\left(\bm{n}\cdot \bm{\chi}_1\right)^2 \bigg[\left(\frac{\nu ^3}{4}-\frac{205 \nu ^2}{28}-\frac{24 \nu }{7}+\left(-\frac{37 \nu ^2}{4}+\frac{64 \nu }{7}+\frac{24}{7}\right) X_1\right) \tilde{C}_{\left.1(\text{ES}^2\right)} -\frac{109 \nu ^4}{12}+\frac{113 \nu ^3}{4}-\frac{515 \nu ^2}{12} \nonumber\\
&\qquad\qquad\quad
+\left(-\frac{13 \nu ^3}{2}-\frac{374 \nu ^2}{3}+\frac{149 \nu }{3}\right) X_1\bigg] 
+ \frac{1}{2}\bm{n}\cdot \bm{\chi}_1 \, \bm{n}\cdot \bm{\chi}_2\left(-\frac{109 \nu ^4}{6}+\frac{133 \nu ^3}{12}-\frac{1301 \nu ^2}{48}\right)
\bigg\rbrace
 + 1 \leftrightarrow 2, \\
%%%%%%%%%%%%%%%%%%%%
B_p^{nS} &= 
\frac{1}{\hat{r}^3} \bigg\lbrace
\left(\bm{n}\cdot \bm{\chi}_1\right)^2 \left[\left(3 \nu ^2-3 \nu +(3-3 \nu ) X_1\right) \tilde{C}_{\left.1(\text{ES}^2\right)}+\frac{5 \nu ^3}{4}-\frac{29 \nu ^2}{4}+\left(\frac{13 \nu }{2}-\frac{31 \nu ^2}{4}\right) X_1\right] \nonumber\\
&\quad\qquad
+\frac{1}{2}\bm{n}\cdot \bm{\chi}_1 \, \bm{n}\cdot \bm{\chi}_2 \left(\frac{5 \nu ^3}{2}+\frac{15 \nu ^2}{4}\right) 
\bigg\rbrace \nonumber\\
&\quad
+\frac{1}{\hat{r}^4} \bigg\lbrace
\left(\bm{n}\cdot \bm{\chi}_1\right)^2 \bigg[\left(-\frac{\nu ^3}{4}+\frac{59 \nu ^2}{2}-\frac{15 \nu }{2}+\left(\frac{37 \nu ^2}{4}-\frac{169 \nu }{4}+\frac{15}{2}\right) X_1\right) \tilde{C}_{\left.1(\text{ES}^2\right)}-\frac{31 \nu ^4}{16}-\frac{413 \nu ^3}{48}-\frac{101 \nu ^2}{24} \nonumber\\
&\qquad\qquad\quad
+\left(\frac{13 \nu ^3}{24}+\frac{763 \nu ^2}{48}+\frac{155 \nu }{24}\right) X_1\bigg] 
+\frac{1}{2}\bm{n}\cdot \bm{\chi}_1 \, \bm{n}\cdot \bm{\chi}_2 \left(-\frac{31 \nu ^4}{8}+\frac{217 \nu ^3}{6}+\frac{1493 \nu ^2}{48}\right) 
\bigg\rbrace
 + 1 \leftrightarrow 2,
\\
%%%%%%%%%%%%%%%%%%%%
B_{np}^{nS} &=
\frac{1}{\hat{r}^3} 
\bigg\lbrace
\left(\bm{n}\cdot \bm{\chi}_1\right)^2\left(-\frac{15 \nu ^3}{4}+15 \nu ^2+\left(\frac{75 \nu ^2}{4}-15 \nu \right) X_1\right)
+\frac{1}{2}\bm{n}\cdot \bm{\chi}_1 \, \bm{n}\cdot \bm{\chi}_2 \left(-\frac{15 \nu ^3}{2}-\frac{45 \nu ^2}{4}\right) 
\bigg\rbrace
\nonumber\\
&\quad 
+\frac{1}{\hat{r}^4}
\bigg\lbrace
\left(\bm{n}\cdot \bm{\chi}_1\right)^2 \bigg[\left(\nu ^3-\frac{31 \nu ^2}{4}+\frac{9 \nu }{2}+\left(-\nu ^2+\frac{121 \nu }{4}-\frac{9}{2}\right) X_1\right) \tilde{C}_{\left.1(\text{ES}^2\right)}-\frac{63 \nu ^4}{16}+\frac{2615 \nu ^3}{48}-\frac{899 \nu ^2}{24} \nonumber\\
&\qquad\qquad\quad
+\left(\frac{509 \nu ^3}{24}-\frac{1331 \nu ^2}{16}+\frac{1223 \nu }{24}\right) X_1\bigg]
+\frac{1}{2}\bm{n}\cdot \bm{\chi}_1 \, \bm{n}\cdot \bm{\chi}_2 \left(-\frac{63 \nu ^4}{8}-\frac{455 \nu ^3}{12}+\frac{63 \nu ^2}{4}\right) 
\bigg\rbrace
 + 1 \leftrightarrow 2,
\\
%%%%%%%%%%%%%%%%%%%%
B_{np}^{SS} &=
\frac{1}{\hat{r}^3} 
\bm{\chi}_1^2 \left[\left(-3 \nu ^2+3 \nu +(3 \nu -3) X_1\right) \tilde{C}_{\left.1(\text{ES}^2\right)}+\frac{9 \nu ^2}{4}+\left(\frac{3 \nu ^2}{2}-\frac{3 \nu }{2}\right) X_1\right] \nonumber\\
&\quad
+\frac{1}{\hat{r}^4} \bigg\lbrace
\bm{\chi}_1^2 \bigg[\left(-\frac{135 \nu ^2}{4}+\frac{11 \nu }{2}+\left(-12 \nu ^2+\frac{173 \nu }{4}-\frac{11}{2}\right) X_1\right) \tilde{C}_{\left.1(\text{ES}^2\right)}+\frac{187 \nu ^4}{48}-\frac{409 \nu ^3}{48}+\frac{781 \nu ^2}{72} \nonumber\\
&\qquad\qquad\quad
+\left(-\frac{187 \nu ^3}{24}+\frac{605 \nu ^2}{144}-\frac{1807 \nu }{72}\right) X_1\bigg] 
+\frac{1}{2}\bm{\chi}_1\cdot \bm{\chi}_2 \left(\frac{187 \nu ^4}{24}-\frac{181 \nu ^3}{4}-\frac{1213 \nu ^2}{18}\right)  
\bigg\rbrace 
+ 1 \leftrightarrow 2,
\\
%%%%%%%%%%%%%%%%%%%%
Q^{S^2} &= 
\frac{\hat{p}_r^3}{\hat{r}^3}
\bigg\lbrace
\bm{n}\cdot \bm{\chi}_1 \hat{\bm{p}}\cdot \bm{\chi}_1 \bigg[\left(20 \nu ^3-35 \nu ^2+\left(35 \nu -20 \nu ^2\right) X_1\right) \tilde{C}_{\left.1(\text{ES}^2\right)}+\frac{45 \nu ^4}{8}-\frac{425 \nu ^3}{12}+\frac{35 \nu ^2}{6} \nonumber\\
&\qquad\qquad\quad
+\left(-\frac{100 \nu ^3}{3}+\frac{145 \nu ^2}{4}-\frac{35 \nu }{6}\right) X_1\bigg]
+ \bm{n}\cdot \bm{\chi}_1 \hat{\bm{p}}\cdot \bm{\chi}_2 \left[\frac{45 \nu ^4}{8}+\frac{10 \nu ^3}{3}-\frac{25 \nu ^2}{16}\right] 
\bigg\rbrace \nonumber\\
&\quad 
+\frac{\hat{p}_r^4}{\hat{r}^3}
\bigg\lbrace
\left(\bm{n}\cdot \bm{\chi}_1\right)^2 \bigg[\left(-35 \nu ^3+\frac{245 \nu ^2}{4}+\left(35 \nu ^2-\frac{245 \nu }{4}\right) X_1\right) \tilde{C}_{\left.1(\text{ES}^2\right)}-\frac{105 \nu ^4}{16}+\frac{2135 \nu ^3}{48}-\frac{35 \nu ^2}{24} \nonumber\\
&\qquad\qquad\quad
+\left(\frac{1085 \nu ^3}{24}-\frac{595 \nu ^2}{16}+\frac{35 \nu }{24}\right) X_1\bigg]
+ \frac{1}{2} \bm{n}\cdot \bm{\chi}_1 \, \bm{n}\cdot \bm{\chi}_2\left[-\frac{105 \nu ^4}{8}-\frac{245 \nu ^3}{12}+\frac{35 \nu ^2}{8}\right]  \nonumber\\
&\quad\qquad
+\bm{\chi}_1^2 \bigg[\left(5 \nu ^3-\frac{35 \nu ^2}{4}+\left(\frac{35 \nu }{4}-5 \nu ^2\right) X_1\right) \tilde{C}_{\left.1(\text{ES}^2\right)}+\frac{5 \nu ^4}{16}-\frac{145 \nu ^3}{48}-\frac{35 \nu ^2}{24} \nonumber\\
&\qquad\qquad\quad
+\left(-\frac{95 \nu ^3}{24}+\frac{5 \nu ^2}{16}+\frac{35 \nu }{24}\right) X_1\bigg] 
+\frac{1}{2}\bm{\chi}_1\cdot \bm{\chi}_2 \left[\frac{5 \nu ^4}{8}+\frac{55 \nu ^3}{12}-\frac{5 \nu ^2}{12}\right] 
\bigg\rbrace
+ 1 \leftrightarrow 2.
\end{align}
\end{subequations}

The spin-cubed corrections in Eq.~\eqref{HoddBB} are
\begin{subequations}
\begin{align}
G_{S^3} &= 
\frac{1}{\hat{r}} \bigg\lbrace
\left(\bm{n}\cdot \bm{\chi}_1\right)^2 \bigg[\left(5 \nu +(5 \nu -5) X_1\right) \tilde{C}_{\left.1(\text{BS}^3\right)}+\left(-\frac{3 \nu ^3}{2}-\frac{9 \nu ^2}{4}+\left(3 \nu ^2+\frac{9 \nu }{4}\right) X_1\right) \tilde{C}_{\left.1(\text{ES}^2\right)}-\nu ^4+7 \nu ^3+\frac{5 \nu ^2}{4} \nonumber\\
&\qquad\qquad
+\left(3 \nu ^3-7 \nu ^2-\frac{5 \nu }{4}\right) X_1\bigg]
+\bm{\chi}_1^2 \left[\left((1-\nu ) X_1-\nu \right) \tilde{C}_{\left.1(\text{BS}^3\right)}+\left(\frac{3 \nu ^2}{4}-\frac{3 \nu  X_1}{4}\right) \tilde{C}_{\left.1(\text{ES}^2\right)}-\frac{5 \nu ^3}{4}+\frac{3 \nu ^2 X_1}{2}\right] \nonumber\\
&\quad\qquad
+\bm{n}\cdot \bm{\chi}_1 \, \bm{n}\cdot \bm{\chi}_2 \left[\left(-\frac{7 \nu ^3}{2}-\frac{5 \nu ^2}{2}-\frac{15 \nu  X_1}{2}\right) \tilde{C}_{\left.1(\text{ES}^2\right)}-2 \nu ^4+\frac{37 \nu ^3}{6}-5 \nu ^2+\left(\frac{\nu ^3}{3}-\frac{11 \nu ^2}{3}\right) X_1\right] \nonumber\\
&\quad\qquad
+\bm{\chi}_1\cdot \bm{\chi}_2 \left[\left(\nu ^3+\frac{\nu ^2}{2}+\frac{3 \nu  X_1}{2}\right) \tilde{C}_{\left.1(\text{ES}^2\right)}+\frac{17 \nu ^3}{12}+\nu ^2+\left(\frac{5 \nu ^3}{6}-\frac{7 \nu ^2}{6}\right) X_1\right] \nonumber\\
&\quad\qquad
+\left(\bm{n}\cdot \bm{\chi}_2\right)^2 \left[\left(2 \nu ^3-\frac{5 \nu ^2}{4}-\frac{15 \nu  X_2}{4}\right) \tilde{C}_{\left.2(\text{ES}^2\right)}-\nu ^4+\frac{89 \nu ^3}{6}-\frac{5 \nu ^2}{2}+\left(\frac{8 \nu ^3}{3}-\frac{25 \nu ^2}{3}\right) X_2\right] \nonumber\\
&\quad\qquad
+\bm{\chi}_2^2 \left[\left(-\nu ^3+\frac{\nu ^2}{4}+\frac{3 \nu  X_2}{4}\right) \tilde{C}_{\left.2(\text{ES}^2\right)}-\frac{31 \nu ^3}{6}+\frac{\nu ^2}{2}+\left(\frac{8 \nu ^2}{3}-\frac{5 \nu ^3}{6}\right) X_2\right]
\bigg\rbrace \nonumber\\
&\quad
+\frac{\hat{L}^2}{\hat{r}^2}
\bigg[
\left(\bm{n}\cdot \bm{\chi}_1\right)^2 \left(\frac{\nu ^4}{2}-\frac{3 \nu ^3}{2}+\frac{\nu ^2}{2}+\left(-\frac{3 \nu ^3}{2}+2 \nu ^2-\frac{\nu }{2}\right) X_1\right) 
+\bm{n}\cdot \bm{\chi}_1 \, \bm{n}\cdot \bm{\chi}_2\left(\nu ^4-\nu ^3 X_1\right)  \nonumber\\
&\qquad\qquad\qquad
+\left(\bm{n}\cdot \bm{\chi}_2\right)^2\left(\frac{\nu ^4}{2}-\frac{\nu ^3 X_2}{2}\right) 
\bigg] \nonumber\\
&\quad
+\hat{p}_r^2
\bigg[
\left(\bm{n}\cdot \bm{\chi}_1\right)^2 \left(-\frac{5 \nu ^4}{2}+\frac{15 \nu ^3}{2}-\frac{5 \nu ^2}{2}+\left(\frac{15 \nu ^3}{2}-10 \nu ^2+\frac{5 \nu }{2}\right) X_1\right) 
+\bm{n}\cdot \bm{\chi}_1 \, \bm{n}\cdot \bm{\chi}_2\left(5 \nu ^3 X_1-5 \nu ^4\right)  \nonumber\\
&\qquad\qquad\qquad
+\left(\bm{n}\cdot \bm{\chi}_2\right)^2\left(\frac{5 \nu ^3 X_2}{2}-\frac{5 \nu ^4}{2}\right) 
\bigg],\\
%%%%%%%%%%%%%%%%%%
\tilde{G}_{S^3} &= G_{S^3} \text{ with }  1 \leftrightarrow 2.
\end{align}
\end{subequations}

The spin-quartic corrections in Eq.~\eqref{HBBansatz} read
\begin{subequations}
\begin{align}
A^{S^4} &= \frac{1}{\hat{r}^5} \bigg\lbrace
\left(\bm{n}\cdot \bm{\chi}_1\right)^4 \bigg[-\frac{63 \nu ^4}{4}+36 \nu ^3+3 \nu ^2+\left(-\frac{15 \nu ^3}{2}+\frac{45 \nu ^2}{2}-\frac{15 \nu }{2}+\left(\frac{45 \nu ^2}{2}-30 \nu +\frac{15}{2}\right) X_1\right) \tilde{C}_{\left.1(\text{ES}^2\right)} \nonumber\\
&\qquad\qquad\qquad
+\left(46 \nu ^3-33 \nu ^2-3 \nu \right) X_1 
+\left(-\frac{35 \nu ^2}{4}+\frac{35 \nu }{4}+\left(\frac{35 \nu }{2}-\frac{35}{4}\right) X_1\right) \tilde{C}_{\left.1(\text{ES}^4\right)}\bigg] \nonumber\\
&\quad\qquad
+\bm{n}\cdot \bm{\chi}_2 \left(\bm{n}\cdot \bm{\chi}_1\right)^3 \left[-3 \nu ^4+3 \nu ^3+\left(6 \nu ^3-6 \nu ^2\right) X_1+\left(3 \nu ^2-3 \nu  X_1\right) \tilde{C}_{\left.1(\text{BS}^3\right)}+\left(3 \nu ^2 X_1-3 \nu ^3\right) \tilde{C}_{\left.1(\text{ES}^2\right)}\right] \nonumber\\
&\quad\qquad
+\bm{\chi}_1^2 \left(\bm{n}\cdot \bm{\chi}_1\right)^2  \bigg[\frac{21 \nu ^4}{2}-27 \nu ^3-3 \nu ^2+\left(9 \nu ^3-27 \nu ^2+9 \nu +\left(-27 \nu ^2+36 \nu -9\right) X_1\right) \tilde{C}_{\left.1(\text{ES}^2\right)} \nonumber\\
&\qquad\qquad\qquad
+\left(-33 \nu ^3+24 \nu ^2+3 \nu \right) X_1
+\left(\frac{15 \nu ^2}{2}-\frac{15 \nu }{2}+\left(\frac{15}{2}-15 \nu \right) X_1\right) \tilde{C}_{\left.1(\text{ES}^4\right)}\bigg] \nonumber\\
&\quad\qquad
+\bm{\chi}_2^2 \left(\bm{n}\cdot \bm{\chi}_1\right)^2 \bigg[-63 \nu ^4+58 \nu ^3+\left(35 \nu ^2-35 \nu  X_2\right) \tilde{C}_{\left.2(\text{BS}^3\right)} \nonumber\\
&\qquad\qquad\qquad
+\left(104 \nu ^3-93 \nu ^2\right) X_2
+\left(\left(6 \nu ^3+12 \nu ^2\right) X_2-12 \nu ^3\right) \tilde{C}_{\left.2(\text{ES}^2\right)}\bigg] \nonumber\\
&\quad\qquad
+\bm{\chi}_1\cdot \bm{\chi}_2 \left(\bm{n}\cdot \bm{\chi}_1\right)^2 \bigg[-\frac{189 \nu ^4}{2}+54 \nu ^3-42 \nu ^2+\left(-\frac{21 \nu ^3}{2}-\frac{105 \nu ^2}{2}+\left(\frac{21 \nu ^2}{2}-6 \nu ^3\right) X_1\right) \tilde{C}_{\left.2(\text{ES}^2\right)} \nonumber\\
&\qquad\qquad\qquad
+\tilde{C}_{\left.1(\text{ES}^2\right)} \left(-\frac{33 \nu ^3}{2}-\frac{105}{2} \tilde{C}_{\left.2(\text{ES}^2\right)} \nu ^2-42 \nu ^2+\left(6 \nu ^3-\frac{21 \nu ^2}{2}\right) X_1\right)\bigg] \nonumber\\
&\quad\qquad
+\frac{1}{2}\left(\bm{n}\cdot \bm{\chi}_2\right)^2 \left(\bm{n}\cdot \bm{\chi}_1\right)^2 \bigg[\frac{21 \nu ^4}{2}-27 \nu ^3-3 \nu ^2+\left(9 \nu ^3-27 \nu ^2+9 \nu +\left(-27 \nu ^2+36 \nu -9\right) X_2\right) \tilde{C}_{\left.2(\text{ES}^2\right)} \nonumber\\
&\qquad\qquad\qquad
+\left(-33 \nu ^3+24 \nu ^2+3 \nu \right) X_2
+\left(\frac{15 \nu ^2}{2}-\frac{15 \nu }{2}+\left(\frac{15}{2}-15 \nu \right) X_2\right) \tilde{C}_{\left.2(\text{ES}^4\right)}\bigg] \nonumber\\
&\quad\qquad 
+\bm{n}\cdot \bm{\chi}_2 \, \bm{\chi}_1^2 \bm{n}\cdot \bm{\chi}_1 \bigg[\frac{23 \nu ^4}{2}-\frac{15 \nu ^3}{2}+\frac{9 \nu ^2}{2}+\left(-2 \nu ^4-6 \nu ^3+\frac{9 \nu ^2}{2}\right) X_1+\left(\frac{3 \nu ^3}{2}-\frac{3 X_1 \nu ^2}{2}+\frac{15 \nu ^2}{2}\right) \tilde{C}_{\left.2(\text{ES}^2\right)} \nonumber\\
&\qquad\qquad\qquad
+\tilde{C}_{\left.1(\text{ES}^2\right)} \left(\frac{9 \nu ^3}{2}+\frac{9 X_1 \nu ^2}{2}+\frac{15}{2} \tilde{C}_{\left.2(\text{ES}^2\right)} \nu ^2+3 \nu ^2\right)\bigg] \nonumber\\
&\quad\qquad
+\frac{1}{2}\bm{n}\cdot \bm{\chi}_2 \, \bm{\chi}_1\cdot \bm{\chi}_2 \, \bm{n}\cdot \bm{\chi}_1 \bigg[\frac{23 \nu ^4}{2}-\frac{15 \nu ^3}{2}+\frac{9 \nu ^2}{2}+\left(\frac{9 \nu ^2}{2}-2 \nu ^4-6 \nu ^3\right) X_2+\left(\frac{9 \nu ^3}{2}+\frac{9 X_2 \nu ^2}{2}+3 \nu ^2\right) \tilde{C}_{\left.2(\text{ES}^2\right)} \nonumber\\
&\qquad\qquad\qquad
+\tilde{C}_{\left.1(\text{ES}^2\right)} \left(\frac{3 \nu ^3}{2}+\frac{15}{2} \tilde{C}_{\left.2(\text{ES}^2\right)} \nu ^2-\frac{3 X_2 \nu ^2}{2}+\frac{15 \nu ^2}{2}\right)\bigg] \nonumber\\
&\quad\qquad
+\bm{\chi}_1^4 \bigg[-\frac{3 \nu ^4}{4}+3 \nu ^3+\left(3 \nu ^3-3 \nu ^2\right) X_1+\left(-\frac{3 \nu ^3}{2}+\frac{9 \nu ^2}{2}-\frac{3 \nu }{2}+\left(\frac{9 \nu ^2}{2}-6 \nu +\frac{3}{2}\right) X_1\right) \tilde{C}_{\left.1(\text{ES}^2\right)} \nonumber\\
&\qquad\qquad\qquad
+\left(-\frac{3 \nu ^2}{4}+\frac{3 \nu }{4}+\left(\frac{3 \nu }{2}-\frac{3}{4}\right) X_1\right) \tilde{C}_{\left.1(\text{ES}^4\right)}\bigg] \nonumber\\
&\quad\qquad
+\frac{1}{2}\left(\bm{\chi}_1\cdot \bm{\chi}_2\right)^2 \left[-3 \nu ^4+3 \nu ^3+\left(6 \nu ^3-6 \nu ^2\right) X_2+\left(3 \nu ^2-3 \nu  X_2\right) \tilde{C}_{\left.2(\text{BS}^3\right)}+\left(3 \nu ^2 X_2-3 \nu ^3\right) \tilde{C}_{\left.2(\text{ES}^2\right)}\right] \nonumber\\
&\quad\qquad
+\frac{1}{2}\bm{\chi}_1^2 \bm{\chi}_2^2 \bigg[18 \nu ^4-\frac{55 \nu ^3}{2}+\left(-2 \nu ^4-47 \nu ^3+\frac{85 \nu ^2}{2}\right) X_2+\left(15 \nu  X_2-15 \nu ^2\right) \tilde{C}_{\left.2(\text{BS}^3\right)} \nonumber\\
&\qquad\qquad\qquad
+\left(9 \nu ^3+\left(-6 \nu ^3-9 \nu ^2\right) X_2\right) \tilde{C}_{\left.2(\text{ES}^2\right)}\bigg] \nonumber\\
&\quad\qquad
+\bm{\chi}_1\cdot \bm{\chi}_2 \, \bm{\chi}_1^2 \bigg[-\frac{3 \nu ^4}{2}+3 \nu ^3-\frac{3 \nu ^2}{2}+\left(-\frac{3 \nu ^3}{2}+\frac{3 X_1 \nu ^2}{2}-\frac{3 \nu ^2}{2}\right) \tilde{C}_{\left.2(\text{ES}^2\right)} \nonumber\\
&\qquad\qquad\qquad
+\tilde{C}_{\left.1(\text{ES}^2\right)} \left(-\frac{3 \nu ^3}{2}-\frac{3}{2} \tilde{C}_{\left.2(\text{ES}^2\right)} \nu ^2-\frac{3 X_1 \nu ^2}{2}\right)\bigg] 
\bigg\rbrace 
+  1 \leftrightarrow 2,\\
%%%%%%%%%%%%%%%%%%%%
Q^{S^4} &= \frac{\hat{p}_r^2}{\hat{r}^4} \Big\lbrace
\bm{\chi}_2^2 \left(\bm{n}\cdot \bm{\chi}_1\right)^2\left(3 \nu ^4-6 \nu ^4 X_2\right)  +\bm{\chi}_1\cdot \bm{\chi}_2 \left(\bm{n}\cdot \bm{\chi}_1\right)^2 \left[9 \nu ^4-3 \nu ^3+\left(6 \nu ^4-12 \nu ^3+3 \nu ^2\right) X_1\right]  \nonumber\\
&\qquad\qquad
+\bm{\chi}_1^2 \bm{n}\cdot \bm{\chi}_1 \, \bm{n}\cdot \bm{\chi}_2 \left[-9 \nu ^4+3 \nu ^3+\left(-6 \nu ^4+12 \nu ^3-3 \nu ^2\right) X_1\right] 
\Big\rbrace \nonumber\\
&\quad
+\frac{\hat{p}_r}{\hat{r}^4} \bigg\lbrace
\hat{\bm{p}}\cdot \bm{\chi}_1 \bigg[
\left(\bm{n}\cdot \bm{\chi}_2\right)^3 \left(-12 \nu ^4+4 \nu ^3+\left(-8 \nu ^4+16 \nu ^3-4 \nu ^2\right) X_2\right)
+\bm{n}\cdot \bm{\chi}_1\bm{\chi}_2^2 \left(8 \nu ^4 X_2-4 \nu ^4\right) \nonumber\\
&\qquad\qquad
+\bm{\chi}_2^2 \bm{n}\cdot \bm{\chi}_2 \left(12 \nu ^4-4 \nu ^3+\left(8 \nu ^4-16 \nu ^3+4 \nu ^2\right) X_2\right)
+ \bm{n}\cdot \bm{\chi}_1 \left(\bm{n}\cdot \bm{\chi}_2\right)^2\left(8 \nu ^4-16 \nu ^4 X_2\right) \nonumber\\
&\qquad\qquad
+\left(\bm{n}\cdot \bm{\chi}_1\right)^2 \bm{n}\cdot \bm{\chi}_2 \left(20 \nu ^4-20 \nu ^3+4 \nu ^2+\left(-8 \nu ^4+16 \nu ^3-4 \nu ^2\right) X_2\right)  \nonumber\\
&\qquad\qquad
+\bm{n}\cdot \bm{\chi}_1\bm{\chi}_1\cdot \bm{\chi}_2 \left(-20 \nu ^4+20 \nu ^3-4 \nu ^2+\left(8 \nu ^4-16 \nu ^3+4 \nu ^2\right) X_2\right)\bigg]
\bigg\rbrace +  1 \leftrightarrow 2.
\end{align}
\end{subequations}
\end{widetext}

% to generate .bbl file, uncomment this and run make script
%\bibliography{inspire}

% run: cp ?.bbl ?_refs.tex    (done in make script)
\input{paper_refs}

\end{document}